\documentclass[%
 reprint,
 superscriptaddress,
 amsmath,amssymb,
 aps,
 prl,
]{revtex4-2}
\usepackage[normalem]{ulem} 
\usepackage{graphicx}
\usepackage{dcolumn}
\usepackage{bm}

\usepackage[dvipsnames]{xcolor}
\definecolor{LC}{rgb}{0.4,0.3,0.9}

\usepackage{comment}
\usepackage{braket}

\newsavebox{\revdisplaybox}

\begin{document}

\preprint{APS/123-QED}

\title{Fault-Tolerant Heisenberg-Limited Quantum Metrology\\with Bosonic Probes}

\author{Daniel G. Li}
\affiliation{Montgomery Blair High School, Silver Spring, MD}
\affiliation{Joint Quantum Institute, University of Maryland, College Park, MD}
\affiliation{Joint Center for Quantum Information and Computer Science, University of Maryland, College Park, MD}
\author{Lorcán O Conlon}
\author{Yu-Xin Wang}
\affiliation{Joint Center for Quantum Information and Computer Science, University of Maryland, College Park, MD}
\author{Erfan Abbasgholinejad}
\author{Twesh Upadhyaya}
\author{Alexey V. Gorshkov}

\affiliation{Joint Quantum Institute, University of Maryland, College Park, MD}
\affiliation{Joint Center for Quantum Information and Computer Science, University of Maryland, College Park, MD}


\begin{abstract}
Quantum sensors have been shown to enable more precise measurements than classical sensors. By using entanglement and coherent evolution, the estimation error can be made to scale better in interrogation time, photon count, and other resources, achieving Heisenberg-limited scaling. However, environmental noise and hardware errors typically destroy the coherence of the sensitive quantum states. This motivates the use of fault-tolerant techniques to create more robust quantum sensing protocols. Such fault-tolerant quantum sensing approaches have been shown to enhance sensitivity for qubit systems, but many practical sensors require bosonic probes. In this paper, we propose a fault-tolerant protocol using bosons to estimate displacements in the presence of
dephasing noise. We construct an optimal error-correcting code under a mean photon-number constraint. Controlled displacements couple the probe to a repetition-encoded qubit register, while repeated boson-parity measurements and resets remove probe noise without erasing the stored logical phase. Under circuit-level bit-flip noise, a distance-$d$ code suppresses logical failures to order $p^{(d+1)/2}$. We show that, for a decay rate $\gamma_A$, a total sensing time $T$ and a mean photon number budget $N$, and for sufficiently small residual signals, the Fisher information scales as $NT^2$ over interrogation times of order $\gamma_A^{-(d+1)/2}$. This is a significant improvement over the non-fault-tolerant approach that achieves Heisenberg scaling only for a total time $\mathcal O(\gamma_A^{-1})$.
\end{abstract}

\maketitle

Quantum metrology utilizes nonclassical resources, such as squeezing and multi-particle entanglement, to achieve measurement precision bounded by the fundamental Heisenberg limit~\cite{giovannetti2011advances,braunstein1994statistical,caves1981quantum}. For estimating an unknown quantity $\theta$ over a total interrogation time $T$, coherent uninterrupted evolution achieves the Heisenberg limit, where the estimator variance scales as $\mathrm{Var}(\hat\theta) \propto 1/T^2$. In contrast, classical strategies or protocols restricted to independent, fixed-duration measurements repeated sequentially within a total time budget $T$ are constrained by the standard quantum limit (SQL), $\mathrm{Var}(\hat\theta) \propto 1/T$. For many practical quantum sensing applications~\cite{degen2017quantum,pirandola2018advances}, bosonic systems provide natural probes. For such systems, the quantum enhancements can also be achieved with respect to mean photon number $N$, e.g., scaling as $\mathrm{Var}(\hat\theta) \propto 1/N$ for displacement sensing. Although coherent quantum evolution potentially offers unprecedented sensitivities for detecting weak forces~\cite{zhang2024quantum,lawrie2019quantum,aspelmeyer2014cavity}, gravitational variations~\cite{aasi2013enhanced,tse2019quantum}, and electromagnetic fields, environmental decoherence limits the coherent interrogation window and degrades the asymptotic scaling back to the SQL~\cite{demkowicz2012elusive,Huelga1997,dorner2009optimal,Escher2011}.

To overcome this fundamental obstacle, active quantum error correction (QEC) has been adapted from quantum computing to metrology~\cite{kessler2014quantum,ozeri2013heisenberg,Arrad2014,dur2014improved}. In the continuous-time setting, the parameter $\theta$ is encoded via a signal Hamiltonian $H = \theta G$ with generator $G$, while the probe simultaneously undergoes Markovian decoherence described by jump operators $\{L_k\}$. The challenge in error-corrected sensing is suppressing the noise from $\{L_k\}$ without erasing the $G$ signal. Refs.~\cite{zhou2018achieving,demkowicz2017adaptive} established the necessary and sufficient Hamiltonian-not-in-Lindblad-span (HNLS) condition, which dictates that Markovian noise can be actively corrected to restore Heisenberg scaling if and only if $G$ has a nonzero component outside the complex linear span of the identity and the noise operators: $G \notin \mathcal{S} = \mathrm{span}\{I, L_k, L_k^\dag, L_k^\dag L_j, \forall j,k\}$.

However, most QEC metrology protocols make idealized assumptions about physical hardware, assuming error-free initial state preparation, noiseless and instantaneous ancilla manipulations, and perfect projective measurements. Actual circuits are not ideal, and faults (e.g., occurring with rate $\gamma_A$) during state preparation, gates, or readout cycles can rapidly corrupt the sensor. Uncorrected circuit errors degrade the effective Heisenberg-limited interrogation window to $T \propto 1/\gamma_A$, after which the system reverts to SQL precision~\cite{rojkov2022bias}. Recently, fault-tolerant techniques have been integrated into discrete-variable qubit sensing by employing verified state preparation and repeated syndrome cycles: Ref.~\cite{sahu2026achieving} demonstrated fault-tolerant Heisenberg scaling in probe number $N$ for parallel qubit sensing, while Ref.~\cite{conlon2026fault} proposed sequential fault-tolerant sensing that extends the coherent interrogation window to $T \propto 1/\gamma_A^{m+1}$ for a distance-$d$ repetition code correcting $m = \lfloor(d-1)/2\rfloor$ bit-flip errors.

Bosonic probes introduce a further recovery problem: the displacement signal moves the oscillator out of its initial code space during sensing. In this work we address this issue, introducing a fault-tolerant sensing protocol for bosonic probes. We use ancilla qubits and controlled displacements to encode the probe state in an antipodal coherent-state superposition, evolve it under the signal $H = \theta (a+a^\dag)$ for a timestep $\tau$, then uncompute those displacements and measure oscillator parity. The outcome of the parity measurement determines a logical phase correction, after which the oscillator is reset and the qubit register retains the accumulated signal. Repeating this cycle separates the oscillator's short sensing intervals from the longer lifetime of the encoded qubit memory, in a similar manner to Ref.~\cite{zaiser2016enhancing}. We derive the fundamental sensitivity limits of this protocol when accounting for potential faults at every stage.

\paragraph{Problem Setting.} We consider the estimation of an unknown coupling parameter $\theta \in \mathbb{R}$ of a continuous linear displacement Hamiltonian acting on an infinite-dimensional bosonic probe mode $\mathcal{H}_P$~\cite{weedbrook2012gaussian}:
\begin{equation}
\label{eq:hamiltonian}
    H = \theta G = \theta (a + a^\dag),
\end{equation}
where $a$ and $a^\dag$ are the bosonic annihilation and creation operators. In the noiseless setting, given a mean photon number budget of $N$, the fundamental precision limit with which $\theta$ can be estimated is (for large $N$) $\mathrm{Var}(\hat\theta) =1/(16NT^2)$ , see Appendix~\ref{app:code}. In this work, the mode is simultaneously subjected to Markovian photon-number dephasing (phase diffusion) governed by the jump operator $L = \sqrt{\kappa}\, a^\dag a = \sqrt{\kappa}\, n$, with dephasing rate $\kappa$. The signal acts during each sensing interval of duration $\tau$, and we model control gates as ideal operations followed by specified qubit faults.

According to the HNLS framework~\mbox{\cite{zhou2018achieving}}, Heisenberg-limited estimation is achievable under active QEC if and only if the generator $G$ is linearly independent of the noise operator span:
\begin{equation}
    G \notin \mathcal{S} = \mathrm{span}\left\{ I, L, L^\dag, L^\dag L\right\} = \mathrm{span}\left\{ I, n, n^2 \right\}.
\end{equation}
Every operator in the noise span $\mathcal{S}$ is strictly diagonal in the Fock basis $\{\ket{n}\}$, whereas the displacement generator $G = a + a^\dag$ has purely off-diagonal matrix elements ($\ket{n}\bra{n \pm 1}$). Consequently, $G$ has a nonzero component outside $\mathcal{S}$ ($G \notin \mathcal{S}$), fulfilling the HNLS condition~\cite{zhou2018achieving} (see Appendix~\ref{app:code} for details).

\paragraph{Ideal Code Construction.}
Let $\alpha=\sqrt N$, where $N$ is the mean photon number at the start of a sensing interval. Encode the ancilla in an odd-length repetition register, $\ket{0_L}=\ket0^{\otimes d}$ and $\ket{1_L}=\ket1^{\otimes d}$, and define
\begin{equation}
 \ket{C_0}=\ket\alpha_P\ket{0_L},\qquad
 \ket{C_1}=\ket{-\alpha}_P\ket{1_L}.
\end{equation}
With $\Pi_C=\ket{C_0}\bra{C_0}+\ket{C_1}\bra{C_1}$ the code projector, coherent-state moments give
\begin{equation}
 \begin{gathered}
 \Pi_Cn\Pi_C=N\Pi_C,\quad
 \Pi_Cn^2\Pi_C=(N^2+N)\Pi_C,\\
 \Pi_CG\Pi_C=2\sqrt N Z_L.
 \end{gathered}
\end{equation}
The first two identities establish correctability of $\{I,n\}$, while the last retains the displacement signal. The projected signal gap $4\sqrt N$ is maximal among two-dimensional codes satisfying the energy constraint and correcting $L=\sqrt{\kappa}n$ (Appendix~\ref{app:code}).

\paragraph{Ideal Code Evolution and Quantum Fisher Information.}
We first describe a sensing protocol that assumes errors only occur during the sensing channel, i.e.~we assume no errors in state preparation, on the ancilla qubits, in the syndrome measurements, or in the final measurement. This will establish a benchmark that we will attempt to reach later in the manuscript without the above assumptions, using a fault-tolerant sensing protocol. The ideal error-corrected sensing protocol is described in detail in Appendix~\ref{app:cycle}, but we present the essential features below.

It will be useful to first describe the effect of an error on the sensing mode when the unknown parameter is exactly $\theta=0$, to understand how the correction works. Consider the initial state  $\ket{0}_P\otimes(\ket{0_L}+\ket{1_L})$ (throughout this section we drop the normalising factor $1/\sqrt{2}$). Apply the controlled displacement:
\begin{equation}
\label{eq:controldisp}
 W=\prod_{j=1}^d
 \exp\!\left[\frac\alpha d(a^\dag-a)Z_j\right],
\end{equation}
which produces the state $\ket{\alpha}\ket{0_L}+\ket{-\alpha}\ket{1_L}$. Let $K$ denote a Kraus operator corresponding to dephasing (e.g. possibly multiple applications of the jump operator $L=\sqrt{\kappa}n$). Define $\ket{\chi}=D(-\alpha)K\ket{\alpha}$ and the photon-number parity operator as $P=(-1)^n$. As $K$ and $P$ commute, and parity reverses displacement $PD(-\alpha)=D(\alpha)P$, we have $D(\alpha)K\ket{-\alpha}=P\ket{\chi}$. Therefore, if $K$ acts on the above state followed by $W^\dagger$ we obtain the state $\ket{\chi}\ket{0_L}+P\ket{\chi}\ket{1_L}$. Defining 
\begin{equation}
 U_{\rm par}=Q_0\otimes I_R+Q_1\otimes X_R,
 \qquad Q_b=\frac{I+(-1)^bP}{2},
\end{equation}
we can write down the recovery operation. Define the even and odd parity states as  $\ket{\chi_0}=Q_0\ket{\chi}$ and $\ket{\chi_1}=Q_1\ket{\chi}$ such that $\ket{\chi}=\ket{\chi_0}+\ket{\chi_1}$ and $P\ket{\chi}=\ket{\chi_0}-\ket{\chi_1}$. The joint state can therefore be written as
\begin{equation}
    \ket{\chi_0}(\ket{0_L}+\ket{1_L})+\ket{\chi_1}(\ket{0_L}-\ket{1_L})\;.
\end{equation}
Thus, by measuring the parity and applying the corresponding correction, the logical memory remains uncorrupted. When the parity measurement reveals $\ket{\chi_0}$ no correction is required. When the parity measurement reveals $\ket{\chi_1}$ a logical $Z_L$ operation is implemented on the ancilla qubits. This enables phase information to be carried from one sensing cycle to the next even as we reset the oscillator. The setting with a non-zero phase already stored in the memory is presented in Appendix~\ref{app:cycle}. Note that this parity measurement enables us to preserve the phase information without revealing which Kraus operator acted.

It is also instructive to consider a full sensing cycle without any noise and with non-zero $\theta$ to understand how the signal accumulates. Suppose that the logical register already stores a phase $\phi$,
so the initial state is
$\ket0_P\otimes(\ket{0_L}+e^{i\phi}\ket{1_L})$.
After applying $W$, evolution under $H=\theta(a+a^\dagger)$
for time $\tau$ produces (defining $\beta=\theta\tau$)
\begin{equation}
\label{eq:idealevol}
 e^{-i\alpha\beta}\ket{\alpha-i\beta}_P\ket{0_L}
 +e^{i\phi+i\alpha\beta}\ket{-\alpha-i\beta}_P\ket{1_L}.
\end{equation}
Applying $W^\dagger$ then gives
\begin{equation}
 \ket{-i\beta}_P\otimes
 \left(e^{-2i\alpha\beta}\ket{0_L}
 +e^{i\phi+2i\alpha\beta}\ket{1_L}\right).
\end{equation}
Thus, for $\alpha=\sqrt N$, the logical relative phase increases from $\phi$ to $\phi+4\sqrt N\,\theta\tau$.
Since the oscillator and register are now uncorrelated, resetting
the oscillator preserves this phase. Unfortunately, the parity recovery introduces a decoherence effect at nonzero $\theta$.
Even in the absence of errors, the oscillator state $\ket{-i\beta}$
has odd parity with probability
\begin{equation}
 q_{\rm odd}=\frac{1-e^{-2\beta^2}}{2}.
\end{equation}
An odd result triggers a $Z_L$ correction and therefore adds $\pi$
to the logical relative phase. After applying the correction and
discarding the parity record, the logical state is
\begin{equation}
 \rho_L'=(1-q_{\rm odd})\ket{\psi_{\phi'}}\bra{\psi_{\phi'}}
 +q_{\rm odd}Z_L\ket{\psi_{\phi'}}\bra{\psi_{\phi'}}Z_L,
\end{equation}
where
$\ket{\psi_{\phi'}}=(\ket{0_L}+e^{i\phi'}\ket{1_L})/\sqrt2$
and $\phi'=\phi+4\sqrt N\,\theta\tau$.
Consequently, the logical coherence is reduced by
$1-2q_{\rm odd}=e^{-2\theta^2\tau^2}$ per cycle. This is described in detail in Appendix~\ref{app:finite}.

We are now in a position to consider a sensing cycle with both bosonic
noise and nonzero $\theta$. For simplicity, we assume the dephasing occurs
immediately after the sensing interval, before applying $W^\dagger$,
and take the control and recovery operations to be ideal. Thus, the signal and dephasing act consecutively in the calculation below.
Suppose that the logical register already stores a phase $\phi$.
Starting from $\ket0_P\otimes(\ket{0_L}+e^{i\phi}\ket{1_L})$,
we apply $W$ from Eq.~\eqref{eq:controldisp} and evolve under $H=\theta(a+a^\dagger)$ for time $\tau$.
Writing $\beta=\theta\tau$ and taking $\alpha=\sqrt N$ real, the state
becomes that given in Eq.~\eqref{eq:idealevol}. 
Let $\{K_\mu\}$ describe the subsequent number-dephasing
channel, where $K_\mu = n^\mu$,
\begin{equation}
 \label{eq:bosonicfinite:channel}
 \mathcal E(\rho)=\sum_\mu K_\mu\rho K_\mu^\dagger,
 \qquad \sum_\mu K_\mu^\dagger K_\mu=I,
 \qquad [K_\mu,P]=0.
\end{equation}
This includes the no-jump contribution and any number of jumps.
For one Kraus operator $K_\mu$, the state after dephasing is
\begin{equation}
 \label{eq:bosonicfinite:error}
 e^{-i\alpha\beta}K_\mu\ket{\alpha-i\beta}_P\ket{0_L}
 +e^{i\phi+i\alpha\beta}K_\mu\ket{-\alpha-i\beta}_P\ket{1_L}.
\end{equation}
We define the unnormalised oscillator states as
\begin{equation}
\begin{split}
     \ket{\chi_\mu^-}&=D(-\alpha)K_\mu D(\alpha)\ket{-i\beta},
 \\
 \ket{\chi_\mu^+}&=D(-\alpha)K_\mu D(\alpha)\ket{+i\beta}.
\end{split}
 \label{eq:bosonicfinite:chi}
\end{equation}
At $\beta=0$, both reduce to the state
$\ket\chi=D(-\alpha)K_\mu\ket\alpha$ used above.
Using $PD(\alpha)=D(-\alpha)P$ and $[K_\mu,P]=0$, applying
$W^\dagger$ to Eq.~\eqref{eq:bosonicfinite:error} gives, up to a
common phase,
\begin{equation}
 \label{eq:bosonicfinite:decoded}
 \ket{\chi_\mu^-}\ket{0_L}
 +e^{i\phi'}P\ket{\chi_\mu^+}\ket{1_L},
\end{equation}
where $\phi'=\phi+4\sqrt N\,\theta\tau$.
The difference from the zero-signal expression is that the two
oscillator states $\ket{\chi_\mu^-}$ and $\ket{\chi_\mu^+}$ need
not coincide.

Measuring oscillator parity with outcome $b\in\{0,1\}$ projects
this state to
\begin{equation}
 \label{eq:bosonicfinite:projected}
 Q_b\ket{\chi_\mu^-}\ket{0_L}
 +(-1)^b e^{i\phi'}Q_b\ket{\chi_\mu^+}\ket{1_L}.
\end{equation}
Applying $Z_L^b$ removes the known sign and leaves
\begin{equation}
 \label{eq:bosonicfinite:corrected}
 Q_b\ket{\chi_\mu^-}\ket{0_L}
 +e^{i\phi'}Q_b\ket{\chi_\mu^+}\ket{1_L}.
\end{equation}
When $\theta=0$, the oscillator factors out and the original
logical state is recovered exactly. At finite $\theta$, the
oscillator can remain correlated with the logical register.
Resetting it to vacuum retains the reduced logical state but
removes these correlations.

To obtain the logical state when the parity record is discarded,
we restore normalisation, trace out the oscillator, and sum over
both parity outcomes and all Kraus operators. In the basis
$\{\ket{0_L},\ket{1_L}\}$, this gives
\begin{equation}
 \label{eq:bosonicfinite:logical}
 \rho_L'=\frac12
 \begin{pmatrix}
  1 & e^{-i\phi'}\mathcal V\\
  e^{i\phi'}\mathcal V^* & 1
 \end{pmatrix},
 \qquad
 \mathcal V=\sum_\mu
 \langle\chi_\mu^+\vert\chi_\mu^-\rangle.
\end{equation}
Here we used $Q_0+Q_1=I$ to sum over parity outcomes.
The overlap can be evaluated without identifying which dephasing
operator acted:
\begin{equation}
 \label{eq:bosonicfinite:visibility}
 \begin{aligned}
 \mathcal V
 &=\bra{+i\beta}D(-\alpha)
   \left(\sum_\mu K_\mu^\dagger K_\mu\right)
   D(\alpha)\ket{-i\beta}\\
 &=\langle+i\beta\vert-i\beta\rangle
 =e^{-2\beta^2}.
 \end{aligned}
\end{equation}
Thus, for this ordering of signal and noise, the dephasing makes no
additional contribution to the recovered logical state. The
logical phase advances from $\phi$ to
$\phi+4\sqrt N\,\theta\tau$, while its visibility is multiplied by
$e^{-2\theta^2\tau^2}$. The joint state after reset is
$\ket0\bra0_P\otimes\rho_L'$, ready for the next cycle. The visibility loss is the same signal-induced parity-recovery effect
described above. An odd parity outcome can occur without a dephasing
error causing an erroneous logical phase flip. Averaging over the parity outcomes reduces the logical coherence. The exact
cancellation in Eq.~\eqref{eq:bosonicfinite:visibility} uses the
assumption that dephasing acts after the sensing evolution; it does
not establish the same result for noise acting throughout sensing. This setting is discussed in Appendix~\ref{app:finite:exact}.

As we show in Appendix~\ref{app:finite:local}, we can compute the Fisher information from the recovered logical
state. Under the assumptions above, with bosonic dephasing applied
after each sensing interval and ideal control and recovery operations,
$M$ cycles of duration $\tau$ give a total sensing time $T=M\tau$ and (defining $V_T=e^{-2\theta^2T\tau}$)
\begin{equation}
 \label{eq:bosonicfi:state}
 \rho_L(T)=\frac12
 \begin{pmatrix}
  1 & V_T e^{-i(4\sqrt N\,\theta T)}\\
  V_T e^{i(4\sqrt N\,\theta T)} & 1
 \end{pmatrix}.
\end{equation}
Measuring the logical qubit in the logical $Y$ basis gives outcome probabilities
\begin{equation}
 \label{eq:bosonicfi:probabilities}
 p_\pm(\theta)=\frac12\left[1\pm V_T
 \sin(4\sqrt N\,\theta T)\right]
\end{equation}
and the corresponding classical Fisher information is given by
\begin{equation}
 \label{eq:bosonicfi:zero}
 \mathcal F_C(0)=16NT^2.
\end{equation}
Thus the protocol retains Heisenberg scaling in total sensing time
locally around $\theta=0$. The visibility remains close to unity
when $\theta^2T\tau\ll1$. These expressions
include the full endpoint-dephasing channel, parity correction, and
oscillator reset. Their independence
of the dephasing strength follows from the cancellation established above and relies on the stated placement of dephasing after sensing. The Fisher information is per execution of the complete $M$-cycle
experiment; $\nu$ independent executions give $\nu\mathcal F_C$. Note that up until this point the logical ancilla qubit could be replaced by a single physical qubit as we have so far assumed that it does not experience any errors.

\paragraph{Fault-Tolerant Sensing Protocol.} The bosonic QEC code above assumes reliable qubit operations. An
incorrect oscillator-parity result can cause an unnecessary logical
$Z_L$ correction, or omit a required one. Such errors occur in every sensing cycle and can therefore destroy the accumulated coherence. For example, if each parity result is incorrectly determined with probability $q$, while all other operations are
ideal, the final logical measurement has
$\mathcal F_C(0)=16NT^2(1-2q)^{2M}$ after $M$ cycles. Thus a fixed nonzero $q$ limits the interval over which the ideal
$T^2$ scaling is retained, limiting the attainable Heisenberg scaling advantage. In this regime Heisenberg scaling is attainable only for a number of cycles that goes as $M\approx 1/q$. Appendix~\ref{app:faults} derives
this result and distinguishes these repeated recovery errors from
errors confined to initial preparation or final readout.

To account for such imperfections, we now extend the above to consider faults in the qubit register and recovery operations. Following Ref.~\cite{conlon2026fault}, we assume independent qubit $X$ faults at preparation, gate-output, idle and measurement locations, with probability at most $p$. We restrict to $X$ faults on the qubits to match the dephasing-only assumption on the bosonic mode. For the bosonic mode, we assume only photon-number dephasing and we assume probe reset is ideal. Displacement gate faults are modeled as an ideal gate followed by independent dephasing on the bosonic mode and $X$  errors on the qubit mode with probability $p$. An output fault leaves the completed displacement unchanged
but reverses that qubit's sign in a subsequent controlled displacement.
At zero signal, the two oscillator branches remain related by parity;
parity recovery removes the oscillator back-action, while the
repetition decoder corrects the data-qubit bit flip. A flip that occurs
before a contribution to the encoding displacement can also change
the coherent amplitude during sensing and hence the signal gained in
that cycle.

We first prepare the qubit register in $\ket{+_L}=(\ket{0_L}+\ket{1_L})/\sqrt2$. Following the construction in Ref.~\cite{conlon2026fault}, we verify the $X$ eigenvalue of a single seed qubit before using CNOT gates to prepare the GHZ state. This verification is necessary because an $X$ error before the initial Hadamard prepares $\ket-$ instead of $\ket+$. Repeating the seed measurement $d$ times and correcting according to the majority result suppresses this sign error to $O(p^r)$, where $r=(d+1)/2$ (Appendix~\ref{app:prep:seed}). Subsequent CNOT faults can produce several data-qubit bit flips. However, on the initial GHZ state, $X^S$ and its complement $X^{S^c}$ are equivalent because $X_L\ket{+_L}=\ket{+_L}$. We therefore extract the repetition-code checks before sensing and establish the initial error frame as described in Appendix~\ref{app:prep:register}. This gives the verified preparation used below (Appendix~\ref{app:prep}).

The oscillator is then prepared in vacuum and the register is coupled to it using $W$ from Eq.~\eqref{eq:controldisp}. Recall that each controlled displacement is an ideal gate followed by the allowed faults on the respective modes, i.e.~bit flips on the qubit modes and dephasing noise on the bosonic mode. An error on the $j$th qubit does not reverse the displacement already applied; it changes the sign with which qubit $j$ controls a later displacement. For $\ket{\psi_L}=c_0\ket{0_L}+c_1\ket{1_L}$, a flip immediately after $W$, followed by an otherwise ideal $W^\dagger$ at zero signal, gives
\begin{equation}
 \begin{aligned}
 W^\dagger X_jW\ket0_P\ket{\psi_L}
 ={}&c_0\ket{2\alpha/d}_P X_j\ket{0_L}\\
 &+c_1\ket{-2\alpha/d}_P X_j\ket{1_L}.
 \end{aligned}
 \label{eq:ftmain:displacementfault}
\end{equation}
The two oscillator branches still differ by parity, so the recovery derived in Appendix~\ref{app:cycle:recovery} removes this oscillator error. The remaining qubit bit flip is corrected by the repetition code. Appendix~\ref{app:recovery:displacements} extends this argument to arbitrary allowed fault histories and explains how correctable encoding errors can change the magnitude of the signal acquired in one cycle.

After each sensing interval, we apply $W^\dagger$ and measure oscillator parity. Each parity measurement uses a fresh readout qubit in $\ket0_R$, the gate $U_{\rm par}$, and computational-basis readout. Because $U_{\rm par}$ commutes with $X_R$, a readout-qubit bit flip can corrupt the measurement record without changing the oscillator parity. This allows us to re repeat this measurement $d$ times and use the majority result. If one record is wrong with probability at most $q\le c p$, then
\begin{equation}
 p_{\rm par}\le\sum_{k=r}^{d}\binom{d}{k}q^k(1-q)^{d-k}
 =O_d(p^r).
 \label{eq:ftmain:parity}
\end{equation}
Here $O_d$ allows constants that depend on the fixed code distance. Appendix~\ref{app:recovery:parity} gives the measurement circuit and checks faults in its component gates. We apply $Z_1$ if the majority is odd and reset the oscillator to vacuum. The qubit register retains the accumulated phase, as in the ideal cycle of Appendix~\ref{app:cycle:gates}.

Next we measure the checks $Z_jZ_{j+1}$. Each check uses two CNOT gates with the data qubits as controls and a fresh readout qubit as target. This orientation prevents a readout-qubit $X$ fault from spreading to the data. We perform $d$ rounds of checks and decode the complete time-ordered syndrome record. The decoder carries the syndrome history to the next sensing cycle. These checks cannot be decoded by taking an independent majority for each stabilizer, because a data error can change its value during measurement. The circuit and decoder are described in Appendices~\ref{app:recovery:checks} and~\ref{app:recovery:decoder}. We apply the indicated $X$ corrections and begin the next cycle. Faults after the last check are included in the next recovery; we do not assume that every cycle begins with a perfectly clean physical register. Appendix~\ref{app:recovery} gives the combined recovery analysis.

After $M$ cycles, the oscillator has been reset and the final measurement acts only on the qubits. We measure $Y_1$ on the first qubit and $X_j$ on each of the others,
then take the product of their $\pm1$ outcomes. On the code space this measures $Y_L=Y_1X_2\cdots X_d$. It requires only single-qubit basis changes and measurements, so terminal faults cannot spread through the register. They contribute an $O_d(p)$ error in the final sign, rather than an error repeated in every sensing cycle (Appendix~\ref{app:meas:readout}). This distinction from repeated recovery errors is quantified in Appendices~\ref{app:faults:records}, \ref{app:faults:information} and~\ref{app:faults:spam}. The final measurement is analysed in Appendix~\ref{app:meas}, with its likelihood and local estimator in Appendix~\ref{app:meas:estimation}.

To compare directly with the ideal QEC result, we again assume that the dephasing acts after each sensing
interval. Let $d$ be odd, $r=(d+1)/2$, and $T=M\tau$. Averaging over
the circuit faults before computing the Fisher information, the final
logical-$Y$ measurement obeys (see Appendix~\ref{app:ftend} for details)
\begin{align}
 \eta_M&\le b_dp+a_d(M+1)p^r,
 \label{eq:ftend:budget-main}\\
 \mathcal F_C^{\rm end}(0)&\ge
 16NT^2(1-2\eta_M)^2,
 \qquad \eta_M<\tfrac12.
 \label{eq:ftend:local-main}
\end{align}
Here $a_d$ and $b_d$ are independent of $M$ and $p$ for a fixed circuit
schedule. The term $b_dp$ accounts for local changes in visibility
and terminal readout errors, whereas $a_d(M+1)p^r$ bounds accumulated
logical failures, including incorrect parity majorities. In particular,
$p_{\rm par}$ is the probability that the majority of the $d$ parity
records is wrong in one cycle, and its contribution to $\eta_M$ is at
most $M p_{\rm par}$. Appendix~\ref{app:ftend} derives the likelihood,
the finite-signal classical Fisher information (CFI), and this bound. With ideal qubit operations
($p=0$), the protocol recovers exactly
$\mathcal F_C^{\rm end}(0)=16NT^2$, independently of the strength of
the endpoint dephasing. For sufficiently small $p$, choosing
$M=\lfloor c p^{-r}\rfloor$ with sufficiently small fixed $c>0$
retains a nonzero fraction of this ideal value. The guaranteed sensing
window therefore extends to times of order $\tau p^{-(d+1)/2}$,
compared with order $\tau/q$ for the unprotected parity-readout model. If the conditional logical-failure probability in every successful cycle is at most
$p_L=O(p^r)$, the no-failure probability is at least $(1-p_L)^M$;
Appendix~\ref{app:ftend:survival} gives the corresponding CFI bound. These statements concern local
estimation around $\theta=0$; a sufficient interval of nonzero residual
signals is also given in Appendix~\ref{app:ftend:window}.

For more realistic considerations, we now allow photon-number dephasing throughout each sensing interval, as in the problem setting. The calculation above where dephasing acts only after the sensing channel remains a useful special case (Appendix~\ref{app:finite:noiseless}); Appendix~\ref{app:finite} also treats simultaneous signal and dephasing, deriving its channel in Appendix~\ref{app:finite:exact}. It reduces the ideal local signal by the factor
\begin{equation}
 f(\kappa\tau)=\frac{1-e^{-\kappa\tau/2}}{\kappa\tau/2},\qquad f(0)=1.
\end{equation}
Appendix~\ref{app:finite:local} derives the resulting ideal-control Fisher information and bounds the finite-signal visibility loss. Combining the preparation, for sufficiently small $p$, recovery and measurement estimates gives $\eta_M$ as in Eq.~\eqref{eq:ftend:budget-main}. The likelihood calculation in Appendix~\ref{app:ft:likelihood} gives
\begin{equation}
 \mathcal F_C(0)\ge16NT^2f(\kappa\tau)^2(1-2\eta_M)^2,
 \label{eq:ftmain:information}
\end{equation}
provided $\eta_M<1/2$. For dephasing acting after the sensing window, the same bound holds with $f$ replaced by $1$.

At fixed $d$, $N$ and cycle duration, sufficiently small $p$ permits $M$ of order $p^{-r}$ while retaining $NT^2$ scaling. This is a significant improvement over $M\approx 1/p$ when not using a fault-tolerant sensing approach (see Appendix~\ref{app:faults}). Appendix~\ref{app:ft:finite} extends this result to a finite interval $|\theta|\le c/(\sqrt N T)$, with $c$ sufficiently small. This allows a nonzero accumulated phase, although the interval narrows as the interrogation time increases. The fault-tolerant bound, including the dependence on cycle duration and uncorrected phase faults, is collected in Appendix~\ref{app:ft}. Assuming operations at a given location take time $\tau_{\rm loc}$ and Poisson bit-flip rate $\gamma_A$, one may use $p=1-e^{-\gamma_A\tau_{\rm loc}}\simeq\gamma_A\tau_{\rm loc}$ as a bound on the probability of a location fault. The resulting sensing window is of order $\tau p^{-(d+1)/2}$, (Appendix~\ref{app:ft:resources}).

\paragraph{Discussion.}
Several assumptions and implementation challenges remain. First, our protocol relies on ancilla qubits with strongly biased noise, where bit-flip faults dominate and phase-flip errors remain sufficiently suppressed ($d\gamma_Z T \ll 1$) over the entire interrogation window to preserve logical coherence and the HNLS condition. In hardware realizations such as stabilized cat qubits, engineering strong noise bias alongside high-fidelity parity readouts and transversal controlled displacements without inducing ancilla phase flips remains an active experimental challenge~\cite{sivak2023real,campagne2020quantum}. Second, our local precision bounds assume that the single-cycle dephasing factor $\kappa\tau$ and the accumulated finite-signal visibility loss $\mathcal{O}(\theta^2 T\tau)$---arising from unrecorded signal-induced parity flips and phase-diffusion fluctuations---remain small over the total sensing time $T$. In practice, the $d$ repeated oscillator-parity and stabilizer-measurement rounds required in each cycle introduce a recovery latency $\tau_{\rm cyc} - \tau$ that grows with code distance, reducing the effective sensing duty cycle $(\tau/\tau_{\rm cyc})^2$ and exposing the logical memory to additional idle faults. Retaining the parity syndrome history across cycles for joint parameter estimation, or applying adaptive compensating drives to keep the residual signal near the $\theta = 0$ operating point, could further mitigate finite-signal visibility decay.

This protocol lays the groundwork for fault-tolerant bosonic metrology beyond linear displacement sensing~\mbox{\cite{xia2023entanglement,brady2023entanglement}}. Future directions include nonlinear generators satisfying HNLS, such as Kerr interactions $H \propto (a^\dag a)^2$ under single-photon loss $L \propto a$, and phase estimation with $H \propto a^\dag a$ under two-photon loss $L \propto a^2$. Grid-based encodings such as Gottesman-Kitaev-Preskill codes~\mbox{\cite{gottesman2001encoding,brady2024advances}} may also help absorb phase-space translations and reduce syndrome-extraction overhead. Ultimately, mapping these protocols onto practical circuits will be essential for applying Heisenberg-limited sensing to fundamental forces.

\begin{acknowledgments}
\textit{Acknowledgments.}---D.G.L., L.O.C., E.A., T.U., and A.V.G.~were supported in part by ONR MURI, NQVL:QSTD:Design:FTL, ARL (W911NF-24-2-0107), AFOSR MURI,  NSF STAQ program, DARPA SAVaNT ADVENT,  DoE ASCR Quantum Testbed Pathfinder program (award No.~DE-SC0024220), and  NSF QLCI (award No.~OSI-2553574). D.G.L., L.O.C., E.A., T.U., and A.V.G.~also acknowledge support from the U.S.~Department of Energy, Office of Science, National Quantum Information Science Research Centers, Quantum Systems Accelerator (award No.~DE-SCL0000121) and from the U.S.~Department of Energy, Office of Science, Accelerated Research in Quantum Computing, Fundamental Algorithmic Research toward Quantum Utility (FAR-Qu).
    
\end{acknowledgments}

\clearpage
\onecolumngrid
\appendix
\setcounter{secnumdepth}{3}

\section{Energy Constraint and Code Optimality}
\label{app:code}
In this appendix we present the fundamental sensitivity limits for the sensing problem considered. We verify that for the Hamiltonian considered in Eq.~\eqref{eq:hamiltonian}, and the jump operator $L=\sqrt{\kappa}n$ the HNLS condition is satisfied, enabling Heisenberg scaling in time.

Let $G=a+a^\dag$ and $\alpha=\sqrt N>0$. The mean photon-number constraint applies to the probe at the start of each sensing interval. For noiseless evolution, under the Hamiltonian in Eq.~\eqref{eq:hamiltonian}, of an arbitrary pure input, $\mathcal F_Q=4T^2\operatorname{Var}(G)$. For a maximum mean photon number of $N$, the Cauchy--Schwarz inequality gives $|\langle a^2\rangle|\le\sqrt{N(N+1)}$, and hence
\begin{equation}
 \mathcal F_Q\le4T^2[2N+1+2\sqrt{N(N+1)}].
 \label{eq:min:noiseless}
\end{equation}
A squeezed vacuum with its anti-squeezed quadrature along $G$ attains this bound. Thus $16NT^2$ is its leading large-$N$ term, not its exact finite-energy value.

For the code $\ket{C_0}=\ket\alpha\ket{0_L}$, $\ket{C_1}=\ket{-\alpha}\ket{1_L}$, orthogonal ancilla labels and coherent-state moments give
\begin{equation}
 \Pi_Cn\Pi_C=N\Pi_C,\qquad
 \Pi_Cn^2\Pi_C=(N^2+N)\Pi_C,\qquad
 \Pi_CG\Pi_C=2\sqrt N Z_L.
 \label{eq:min:KL}
\end{equation}
The first two identities are the Knill--Laflamme conditions for $\{I,n\}$~\cite{KnillLaflamme1997}. The signal is outside the algebraic Lindblad span $\operatorname{span}\{I,n,n^2\}$ because that span is diagonal in the Fock basis. On the infinite-dimensional oscillator space, these operators are unbounded; the explicit recovery in Appendix~\ref{app:cycle} below establishes the sensing result without invoking a Hilbert--Schmidt projection of $G$ onto this span.

For any two-dimensional code correcting $n$, write $\Pi_Cn\Pi_C=\nu\Pi_C$, where $\nu\le N$. Every normalized code vector obeys $|\langle G\rangle|\le2\sqrt\nu$. The spectral gap of $\Pi_CG\Pi_C$ is therefore at most $4\sqrt N$, giving projected-code QFI at most $16NT^2$. Equality requires extremal eigenvectors of $a\otimes I$ with eigenvalues $\pm\sqrt N$, so their oscillator factors are $\ket{\pm\sqrt N}$. Their ancilla factors must be orthogonal because $\langle\sqrt N|-\sqrt N\rangle=e^{-2N}\ne0$. The displayed code attains the maximal projected signal gap. This statement does not assert optimality of its finite-cycle recovery for arbitrary nonzero signal.

\section{One Ideal-Control Sensing and Recovery Cycle}
\label{app:cycle}
In this appendix we describe one full cycle of the ideal error correcting code. In Appendix~\ref{app:cycle:gates} we describe the encoding, sensing and bosonic mode reset. In Appendix~\ref{app:cycle:recovery} we then verify that when $\theta=0$ this process exactly corrects errors occuring on the bosonic mode.
\subsection{Encoding, sensing and reset}
\label{app:cycle:gates}

Let $d$ be odd, $\ket{0_L}=\ket0^{\otimes d}$, $\ket{1_L}=\ket1^{\otimes d}$ and $\alpha=\sqrt N$. Define
\begin{equation}
 W=\prod_{j=1}^dV_j,\qquad
 V_j=\exp[(\alpha/d)(a^\dag-a)Z_j],\qquad
 P=(-1)^n,\qquad Q_b=\frac{I+(-1)^bP}{2}.
 \label{eq:min:gates}
\end{equation}
For an arbitrary retained logical state,
\begin{equation}
 W\ket0_P(c_0\ket{0_L}+c_1\ket{1_L})
 =c_0\ket\alpha_P\ket{0_L}+c_1\ket{-\alpha}_P\ket{1_L}.
 \label{eq:min:encode}
\end{equation}
The oscillator senses for time $\tau$. In the main text we considered the setting where the dephasing acts after the signal unitary. For simultaneous signal and dephasing, its evolution instead obeys
\begin{equation}
 \dot\rho=-i\theta[G,\rho]+\kappa\mathcal D[n](\rho),\qquad
 \mathcal D[n](\rho)=n\rho n-\tfrac12\{n^2,\rho\}.
 \label{eq:min:master}
\end{equation}
Both cases use the same recovery: apply $W^\dagger$, measure oscillator parity, apply $Z_1$ for an odd outcome, and reset only the oscillator to vacuum. Under ideal controls, ancilla qubits, and measurements, one parity measurement suffices. The parity measurement circuit uses a readout qubit in $\ket0_R$, the gate
\begin{equation}
 U_{\rm par}=Q_0\otimes I_R+Q_1\otimes X_R,
 \label{eq:min:paritygate}
\end{equation}
and computational-basis readout. The repeated measurements and data checks needed for faulty controls are introduced in Appendix~\ref{app:recovery}.

Reset is the channel $\rho_{PL}\mapsto\ket0\bra0_P\otimes\operatorname{Tr}_P\rho_{PL}$. It preserves the reduced logical state even when the oscillator remains correlated with the register. Exact preservation of the input logical coherence additionally requires the recovery identities proved next, or their form for  non-zero $\theta$ in Appendix~\ref{app:finite}.

\subsection{Exact recovery at zero signal}
\label{app:cycle:recovery}

At $\theta=0$, an exact Kraus representation of number dephasing over time $\tau$ is
\begin{equation}
 K_\ell=\frac{(\kappa \tau)^{\ell/2}}{\sqrt{\ell!}}
 e^{-\kappa \tau n^2/2}n^\ell,\qquad \ell=0,1,\ldots.
\end{equation}
These operators commute with parity and obey $\sum_\ell K_\ell^\dagger K_\ell=I$. They include both no-jump evolution and arbitrarily many jumps of the same Lindblad operator $\sqrt\kappa n$.

For each Kraus term, define $\ket\chi=D(-\alpha)K_\ell\ket\alpha$. Parity symmetry gives the state after uncomputation as
\begin{equation}
 c_0\ket\chi\ket{0_L}+c_1P\ket\chi\ket{1_L}.
\end{equation}
For parity outcome $b$,
\begin{equation}
 (Q_b\otimes I)\ket\Psi
 =Q_b\ket\chi\otimes
 [c_0\ket{0_L}+(-1)^b c_1\ket{1_L}].
 \label{eq:min:factor}
\end{equation}
Therefore, applying the correction $Z_1^b=Z_L^b$ on the code restores the logical state. The oscillator then factors out and can be reset. The same operation works for every Kraus term, so the result applies to the full channel, mixed logical inputs and logical inputs entangled with a reference. It is exact for any $\kappa \tau$ at zero signal.

\section{Finite-Signal Evolution and Fisher Information}
\label{app:finite}
In this appendix we discuss the effect of non-zero $\theta$, which introduces an additional decoherence. Appendix~\ref{app:finite:noiseless} explains the origin of this effect. In Appendix~\ref{app:finite:exact} we extend this to the setting where the dephasing and sensing act simultaneously. Finally, in Appendix~\ref{app:finite:local} we compute the CFI in both of these settings.

\subsection{Endpoint dephasing and oscillator reset}
\label{app:finite:noiseless}
We first show how a non-zero $\theta$ introduces an additional decoherence. The reset used above does not require the true signal to vanish. With $\kappa=0$ and $\beta=\theta\tau$, the state immediately after uncomputation is exactly
\begin{equation}
 W^\dag e^{-i\theta\tau G}W\ket0_P\ket{\psi_L}
 =\ket{-i\beta}_P\otimes e^{-2i\alpha\theta\tau Z_L}\ket{\psi_L}.
 \label{eq:min:free}
\end{equation}
Thus the qubits already carry relative phase $4\alpha\theta\tau$. The parity measurement of $\ket{-i\beta}$ gives an odd result with probability $q_{\rm odd}=(1-e^{-2\beta^2})/2$, even in the absence of errors. As there is no way to distinguish whether the signal or an error caused the odd parity measurement result, applying the prescribed odd-outcome correction and discarding the syndrome record gives
\begin{equation}
 \mathcal E_\tau(\rho_L)=
 (1-q_{\rm odd})R_\theta\rho_LR_\theta^\dag
 +q_{\rm odd}Z_LR_\theta\rho_LR_\theta^\dag Z_L,
 \qquad R_\theta=e^{-2i\alpha\theta\tau Z_L}.
 \label{eq:min:noiselesschannel}
\end{equation}
Therefore, an odd outcome produced by the signal consequently reduces the unconditional coherence. The sensitivity bounds derived below use the final qubit measurement alone, after the syndrome records have served their correction purpose. Keeping those records for estimation may supply additional information in principle, however we do not analyse this case here.

Suppose that number dephasing acts after the signal evolution but
before $W^\dagger$. For a particular Kraus operator $K_\mu$,
the oscillator states after correction depend on which parity
result was obtained. Summing over both results and all Kraus
operators, the logical coherence contains
$\sum_\mu\langle\chi_\mu^+|\chi_\mu^-\rangle$.
Since the dephasing channel is trace preserving,
$\sum_\mu K_\mu^\dagger K_\mu=I$, this sum equals
$\langle+i\beta|-i\beta\rangle=e^{-2\beta^2}$,
as shown in Eqs.~\eqref{eq:bosonicfinite:corrected}--
\eqref{eq:bosonicfinite:visibility}. Dephasing may therefore
change the probabilities of the parity outcomes, but it does
not change the recovered logical state after those outcomes
are averaged. Equations~\eqref{eq:min:noiselesschannel}
and~\eqref{eq:min:finitecoherence} also hold for dephasing
of arbitrary strength applied after the signal. When dephasing
acts during sensing, this calculation no longer applies:
$n$ does not commute with the signal generator $G$.

For $\ket{+_L}$ and $M$ cycles, $T=M\tau$, the off-diagonal element is
\begin{equation}
 [\rho_L(T)]_{01}=\tfrac12 e^{-2\theta^2T\tau}e^{-4i\alpha\theta T}
.
 \label{eq:min:finitecoherence}
\end{equation}
The phase is finite and accumulates across resets. At fixed $\theta\ne0$ and fixed $\tau$, however, the decaying visibility eventually dominates implying there is some finite optimal sensing time. Retaining a long coherent run requires $\theta^2T\tau$ to remain small, or a residual signal made small by a compensating known drive. Throughout this work, we shall restrict $\theta$ to be sufficiently small such that this condition is satisfied. Future work will involve constructing an explicit bit-wise learning procedure to remove this restriction. 

\subsection{Exact channel with number dephasing}
\label{app:finite:exact}

We now use the simultaneous Lindblad evolution in Eq.~\eqref{eq:min:master} to extend the above results to the setting where the dephasing and displacement act simultaneously. Assuming ideal controls, we can model one cycle as preserving logical populations and multiplying the logical off-diagonal element by a factor $\chi_\tau(\theta)$. We obtain this factor by averaging unitary evolutions with a randomly rotating oscillator phase. Let $B_s$ be standard real Brownian motion: $B_0=0$, and increments over disjoint intervals are independent Gaussian variables with $\mathbb E(B_t-B_s)=0$ and $\mathbb E[(B_t-B_s)^2]=t-s$ for $t>s$. For each path, let $U_s$ solve the Stratonovich equation
\begin{equation}
 dU_s=-i\theta G U_s\,ds-i\sqrt\kappa n U_s\circ dB_s,
 \qquad U_0=I.
\end{equation}
Averaging $U_s\rho(0)U_s^\dag$ over the Brownian paths gives the density operator governed by Eq.~\eqref{eq:min:master}. To remove the random rotation, define
\begin{equation}
 R_s=e^{-i\sqrt\kappa B_s n},\qquad V_s=R_s^\dag U_s.
\end{equation}
The interaction-picture evolution then satisfies
\begin{equation}
 i\frac{dV_s}{ds}
 =\theta R_s^\dag G R_s V_s
 =\theta\bigl(e^{-i\sqrt\kappa B_s}a
                 +e^{i\sqrt\kappa B_s}a^\dag\bigr)V_s.
\end{equation}
Thus the direction of the signal-induced displacement rotates along each path. Its accumulated complex amplitude is
\begin{equation}
 A_\tau=\int_0^\tau e^{i\sqrt\kappa B_s}\,ds.
\end{equation}
With no phase diffusion, $A_\tau=\tau$. For a general path, the displacements acquired at different times add with different phases. Their commutators contribute only a scalar phase, so
\begin{equation}
 U_\tau=R_\tau D(-i\theta A_\tau)e^{i\theta^2 c[B]},
\end{equation}
where $c[B]$ is real and depends on the Brownian path. The parity-corrected coherence for a fixed path is $\langle-\alpha|U_\tau^\dag P U_\tau|\alpha\rangle$. Since $P$ commutes with $R_\tau$, the random final rotation and the scalar phase cancel from this overlap. Using the coherent-state displacement overlap gives
\begin{equation}
 \chi_\tau(\theta)=\mathbb E_B\left[
 e^{-2\theta^2|A_\tau|^2-4i\alpha\theta\operatorname{Re}A_\tau}
 \right].
 \label{eq:min:chi}
\end{equation}
Here $\mathbb E_B$ averages over all Brownian paths during one sensing interval. Independent intervals have independent paths; the average does not condition on a particular oscillator-parity outcome.

An equivalent deterministic calculation solves
\begin{equation}
 \dot\sigma_\theta=-i\theta\{G,\sigma_\theta\}
       +\kappa\mathcal D[n](\sigma_\theta),\qquad
 \sigma_\theta(0)=\ket\alpha\bra\alpha,\qquad
 \chi_\tau(\theta)=\operatorname{Tr}\sigma_\theta(\tau).
 \label{eq:min:ode}
\end{equation}
No small-$\theta$ approximation has entered either representation. Independent oscillator resets give $[\rho_L]_{01}=\chi_\tau(\theta)^M/2$ for an initial $\ket{+_L}$. Thus the logical phase is $-M\arg\chi_\tau(\theta)$ and its visibility is $|\chi_\tau(\theta)|^M$.

\subsection{Local information and a finite-cycle visibility bound}
\label{app:finite:local}
We now compute the CFI both when dephasing acts after the sensing channel, and when dephasing acts during the sensing channel. For endpoint dephasing, put $v(\theta)=e^{-2\theta^2T\tau}$ and $\Phi=4\sqrt N\,\theta T$. The final logical-$Y$ probabilities are $(1\pm v\sin\Phi)/2$, so
\begin{equation}
 \mathcal F_C^{\rm end}(\theta)=
 \frac{v^2[4\sqrt N\,T\cos\Phi-4\theta T\tau\sin\Phi]^2}
 {1-v^2\sin^2\Phi},\qquad
 \mathcal F_C^{\rm end}(0)=16NT^2.
 \label{eq:finite:endfi}
\end{equation}
This is the finite-signal extension of the main-text endpoint calculation. 

For simultaneous signal and dephasing, the recovered channel instead contains the average in Eq.~\eqref{eq:min:chi}. Since $\mathbb E\operatorname{Re}A_\tau=\tau f(\kappa\tau)$,
\begin{equation}
 f(x)=\frac{1-e^{-x/2}}{x/2},\quad f(0)=1,\qquad
 \chi_\tau(0)=1,\quad
 \chi_\tau'(0)=-4i\sqrt N\tau f(\kappa\tau).
\end{equation}
Logical $Y_L$ readout has mean $y(\theta)=-\operatorname{Im}\chi_\tau(\theta)^M$. Therefore $y(0)=0$, and
\begin{equation}
 \mathcal F_C(0)=16NT^2f(\kappa\tau)^2.
 \label{eq:min:localfi}
\end{equation}
This is an attainable classical Fisher information, evaluated at $\theta=0$. Note that the factor $f(\kappa\tau)^2$ acts as a constant factor reduction in the CFI for fixed $\kappa$ and $\tau$. Taking $M$, and therefore $T$ to be large demonstrates that this CFI achieves the same scaling as the optimal CFI derived in Appendix~\ref{app:code}.

The finite-signal loss can be bounded separately. Put $X=\operatorname{Re}A_\tau$ and $\mu=\mathbb EX$. Using $|A_\tau|\le\tau$ and $|e^{-iz}-1+iz|\le z^2/2$ gives
\begin{equation}
 |\chi_\tau(\theta)-e^{-4i\alpha\theta\mu}|
 \le\delta_\tau:=2\theta^2\tau^2+8N\theta^2\operatorname{Var}(X).
 \label{eq:min:cycleloss}
\end{equation}
Moreover, $1-\cos x\le x^2/2$ and $\mathbb E(B_s^2B_t^2)=st+2\min(s,t)^2$ imply
\begin{equation}
 \operatorname{Var}(X)\le\mathbb E(X-\tau)^2
 \le\frac{7}{48}\kappa^2\tau^4.
\end{equation}
It follows that $|\chi_\tau|^M\ge1-M\delta_\tau$ whenever the right-hand side is positive. A sufficient condition for small visibility loss is
\begin{equation}
 2\theta^2T\tau+\frac76N\kappa^2\theta^2T\tau^3\ll1.
 \label{eq:min:finiteregime}
\end{equation}
If $M=a_0p^{-r}$, this sufficient condition becomes
\begin{equation}
 |\theta|\tau\ll
 \frac{p^{r/2}}{\sqrt{a_0[2+(7/6)N(\kappa\tau)^2]}}.
\end{equation}
It is weaker than the small-accumulated-phase condition used for the fixed-measurement Fisher-information bound below. This is a visibility estimate, not a substitute for differentiating the likelihood. Appendix~\ref{app:ft} gives a Fisher-information bound on a nonzero interval of residual signals, including circuit faults.

\section{Why the Recovery Must Tolerate Faults}
\label{app:faults}
We now demonstrate the necessity of introducing a fault-tolerant quantum sensing protocol. The ideal-control calculation does not by itself establish robustness to faults in recovery. In Appendix~\ref{app:faults:records} we introduce one additional fault
process and show that the advantage of the naive ideal error corrected sensing process is lost. The fault we introduce is that each oscillator-parity measurement returns the wrong bit
with probability $q$, independently between cycles. All other
operations are ideal. This process can arise from an $X$ fault on
the readout qubit immediately before its computational-basis
measurement. Taking $q=p$ and all other qubit fault probabilities
to be zero gives an instance of the stated circuit fault model. For simplicity we retain the bosonic-noise ordering used above: sensing for time
$\tau$ is followed by number dephasing, ideal uncomputation, parity
measurement and correction, and oscillator reset. In Appendix~\ref{app:faults:information} we compute the CFI when this additional fault is introduced. Finally, in Appendix~\ref{app:faults:spam} we demonstrate that introducing faults at other locations, such as state preparation, and the final readout can have a different effect.

\subsection{An incorrect parity result causes a logical phase flip}
\label{app:faults:records}
First, we consider introducing faults in the parity readout. Let $b$ be the actual oscillator-parity outcome and let $e$ indicate whether the reported bit is wrong. The correction applied to the
register is
\begin{equation}
 Z_L^{b\oplus e}=Z_L^e Z_L^b.
\end{equation}
Thus a wrongly reported parity introduces an additional logical $Z_L$, regardless
of the actual parity. Since this additional error is independent of the
state and the true outcome, averaging over it gives the channel
\begin{equation}
 \label{eq:ftmot:recordchannel}
 \mathcal Z_q(\rho_L)=(1-q)\rho_L+qZ_L\rho_LZ_L.
\end{equation}
In particular, $[\rho_L]_{01}$ is multiplied by $1-2q$.
A single readout-qubit $X$ fault can therefore become a logical
phase error through the conditional correction. The repetition
code's $Z_jZ_{j+1}$ checks do not detect this error: $Z_L$, represented
by $Z_1$ on the code space, commutes with every such check.

With accurate parity reporting, one cycle multiplies the logical
coherence by
$e^{-2\theta^2\tau^2}e^{-4i\sqrt N\,\theta\tau}$.
Including Eq.~\eqref{eq:ftmot:recordchannel} instead gives
\begin{equation}
 \label{eq:ftmot:onecycle}
 [\rho_L]_{01}\longmapsto
 (1-2q)e^{-2\theta^2\tau^2}
 e^{-4i\sqrt N\,\theta\tau}[\rho_L]_{01}.
\end{equation}
Starting from $\ket{+_L}=(\ket{0_L}+\ket{1_L})/\sqrt2$, after
$M$ cycles and total sensing time $T=M\tau$, the state is
\begin{equation}
 \label{eq:ftmot:state}
 \rho_L(T)=\frac12
 \begin{pmatrix}
 1 & A_M(\theta)e^{-4i\sqrt N\,\theta T}\\
 A_M(\theta)e^{4i\sqrt N\,\theta T} & 1
 \end{pmatrix},
 \qquad
 A_M(\theta)=(1-2q)^M e^{-2\theta^2T\tau}.
\end{equation}
Here and below $q$ is independent of the unknown parameter.
The factor $(1-2q)^M$ remains even at $\theta=0$, where the ideal
bosonic recovery would preserve the logical state exactly.

\subsection{Fisher information and the useful sensing time}
\label{app:faults:information}
We now compute the CFI when accounting for the above imperfection. A logical $Y_L$ measurement has probabilities
\begin{equation}
 p_\pm(\theta)=\frac12\left[1\pm A_M(\theta)
 \sin(4\sqrt N\,\theta T)\right].
\end{equation}
Writing $\Phi=4\sqrt N\,\theta T$, its classical Fisher information is
\begin{equation}
 \label{eq:ftmot:finiteFI}
 \mathcal F_C(\theta)=
 \frac{A_M(\theta)^2
 \left[4\sqrt N\,T\cos\Phi-4\theta T\tau\sin\Phi\right]^2}
 {1-A_M(\theta)^2\sin^2\Phi}.
\end{equation}
At the operating point $\theta=0$, this reduces to
\begin{equation}
 \label{eq:ftmot:zeroFI}
 \mathcal F_C(0)=16NT^2(1-2q)^{2M}.
\end{equation}
For $0<q<1/2$, the state at this operating point has full rank.
Its Bloch-vector length is $(1-2q)^M$, the derivative of this
length vanishes at zero signal, and its azimuthal angle has derivative
$4\sqrt N\,T$. Consequently, its quantum Fisher information is
also
\begin{equation}
 \mathcal F_Q(0)=16NT^2(1-2q)^{2M}.
\end{equation}
The loss in Eq.~\eqref{eq:ftmot:zeroFI} therefore cannot be repaired
by changing the final logical measurement on this reduced state.

For small $q$, $(1-2q)^{2M}=\exp[-4Mq+O(Mq^2)]$.
Maintaining a visibility factor bounded away from zero requires
$Mq=O(1)$; recovering the ideal prefactor requires $Mq\ll1$.
At fixed nonzero $q$ and fixed $\tau$, the ratio
$\mathcal F_C(0)/(16NT^2)$ vanishes as $M$ increases.
Treating $M$ as continuous, the Fisher information per complete
experiment is maximised at
\begin{equation}
 M_{\rm opt}=-\frac{1}{\log(1-2q)}
 \simeq\frac{1}{2q},
 \qquad T_{\rm opt}=M_{\rm opt}\tau,
\end{equation}
with an integer number of cycles chosen near this value. Therefore, the effect of this additional fault is to limit the time period over which Heisenberg scaling can be achieved. 

\subsection{Preparation and final readout have a different effect}
\label{app:faults:spam}
We now consider alternative faults that occur only once during the entire sensing process. Such errors do not necessarily cause the $M$-dependent decay from above. To illustrate this distinction, suppose the prepared logical state, after any initial decoding, is
\begin{equation}
 \rho_{L,\mathrm{prep}}=\frac12(I+v_{\mathrm{prep}}X_L),
 \qquad 0\le v_{\mathrm{prep}}\le1,
\end{equation}
where $v_{\mathrm{prep}}$ describes its initial contrast. Suppose
also that the final binary readout is independently flipped with
probability $\epsilon_{\mathrm{ro}}$. For this preparation model,
\begin{equation}
 \label{eq:ftmot:SPAM}
 \mathcal F_C(0)=16NT^2
 v_{\mathrm{prep}}^2(1-2\epsilon_{\mathrm{ro}})^2
 (1-2q)^{2M}.
\end{equation}
If $q=0$ and the preparation and final-readout factors are fixed
and nonzero, the Fisher information still scales as $T^2$ for arbitrarily large $T$, in contrast to the case where $q\neq0$. For example, reporting the parity of $d$ final measurement bits,
each independently flipped with probability $p$, multiplies the
readout contrast by $(1-2p)^d$. At fixed $d$, this is independent
of the number of sensing cycles. Note that this does not mean that any physical fault occurring during state preparation or the final measurement does not affect Heisenberg scaling. The effect of specific fault locations must be examined as they may cause correlated errors across multiple qubits at a later stage.

\section{Verified State Preparation}
\label{app:prep}
In this appendix we describe the fault-tolerant state preparation of the ancilla register in the logical state $\ket{+_L}$, which is performed before the oscillator interacts with the signal. Following the verified-seed and repetition-code construction of Ref.~\cite{conlon2026fault}, Appendix B.6, we first show in Appendix~\ref{app:prep:seed} how repeated $X$-basis checks on a single seed qubit suppress phase-flip preparation errors to $O_d(p^r)$. In Appendix~\ref{app:prep:register} we then analyze the subsequent CNOT fan-out and initial stabilizer syndrome extraction used to establish a valid initial error frame for the first sensing cycle. Let $d=2m+1$ and $r=m+1$. All estimates below are at fixed $d$ and fixed circuit schedule, with independent elementary qubit faults of probability at most $p$.

\subsection{Verification of the seed phase}
\label{app:prep:seed}
We first fault-tolerantly prepare the single-qubit seed state $\ket{+}$ and verify that a bit-flip fault before the initial Hadamard gate has not prepared $\ket{-}$ instead, using $d$ repeated nondestructive readout checks to suppress the probability of a wrong seed sign to $O_d(p^r)$.

We initialize the seed in $\ket0$ and apply $H$. A preparation $X$ before $H$ produces $\ket-$, since $HX\ket0=\ket-$. In contrast, an $X$ after $H$ preserves either $X$ eigenstate up to a scalar sign. This is why the seed must be checked before the CNOT fan-out.

For each check, prepare a fresh readout qubit $R$ in $\ket+$, apply a CNOT with $R$ as control and the seed as target, and measure $X_R$ using a Hadamard followed by computational-basis readout. For a seed eigenstate $\ket{x}_S$ with $X_S\ket{x}_S=x\ket{x}_S$, $x=\pm1$, the ideal circuit gives
\begin{equation}
 \operatorname{CNOT}_{R\to S}\ket+_R\ket{x}_S
 =\frac{\ket0_R+x\ket1_R}{\sqrt2}\ket{x}_S.
\end{equation}
Thus the seed eigenvalue is measured without changing it. An $X_S$ fault does not change that eigenvalue. A control $X_R$ propagated through the CNOT can introduce $X_S$, but this too acts as a scalar on the seed. Readout preparation, gate-output and measurement faults can flip the reported sign. Each check therefore measures the same fixed eigenvalue with record-error probability at most $q_s\le c_s p$, where $c_s$ counts the relevant locations.

Repeat the check $d$ times and apply $Z$ to the seed if the majority is negative. A fault after this corrective $Z$ is an $X$ and leaves the corrected $\ket+$ unchanged. The probability of a wrong seed sign is bounded by
\begin{equation}
 \epsilon_{\rm seed}\le\sum_{k=r}^{d}\binom{d}{k}q_s^k(1-q_s)^{d-k}
 =O_d(p^r).
 \label{eq:prep:seed}
\end{equation}
Verification uses only fresh readout qubits and parameter-independent operations, so it can be completed before sensing begins.

\subsection{Encoding the register and its initial syndrome}
\label{app:prep:register}
Next, we expand the verified seed into a $d$-qubit GHZ repetition-code state via CNOT fan-out and perform $d$ rounds of stabilizer checks (using the circuit detailed in Appendix~\ref{app:recovery:checks}) to identify and correct propagated bit-flip patterns before sensing begins.

Use the verified seed and $d-1$ qubits initialized in $\ket0$ to prepare $\ket{+_L}$ by a CNOT fan-out. Under $X$ faults, the output is a GHZ state with a definite bit pattern,
\begin{equation}
 X^S\ket{+_L}=\frac{\ket{x}+\ket{\bar x}}{\sqrt2},
 \qquad X^S=\prod_{j\in S}X_j,
\end{equation}
apart from the seed-sign failure already bounded above. Even a fault that propagates to many data qubits leaves a state of this form. Moreover,
\begin{equation}
 X^S\ket{+_L}=X^{S^c}\ket{+_L},\qquad
 \min\{|S|,|S^c|\}\le m.
\end{equation}
The complete stabilizer syndrome identifies the pair $\{S,S^c\}$; either representative restores the same initial state. This equivalence is specific to preparation of $\ket{+_L}$ and must not be used after an unknown phase has accumulated.

Measure the checks $Z_jZ_{j+1}$ over $d$ rounds using the circuit in Appendix~\ref{app:recovery:checks}. Decode these records with the preparation boundary of the repetition-memory construction, allowing the initial syndrome to be unknown after fan-out. In particular, the decoder must not assume that a propagated preparation fault was a collection of independent data faults. Correct the inferred bit pattern and carry the final syndrome information into the first sensing cycle. Faults during verification can leave residual correctable errors; the preparation guarantee is a valid memory input up to these errors, rather than a fault-free physical register. The seed error and decoded preparation failure are $O_d(p^r)$; the probability of a local residual fault affecting the first encoding is $O_d(p)$, as included in Appendix~\ref{app:recovery:decoder}.

The oscillator is prepared in vacuum after verification. Applying $W$ then produces the first bosonic code state. Faults in this application of $W$ are treated in exactly the same way as faults in later encodings.

\section{Fault-Tolerant Syndrome Extraction and Correction}
\label{app:recovery}

In this appendix we analyze the fault-tolerant syndrome extraction and correction protocols used to suppress dephasing on the bosonic probe and bit-flip faults on the ancilla register. In Appendix~\ref{app:recovery:displacements} we examine how faults during controlled displacements propagate to the oscillator branches and affect the single-cycle signal gain. In Appendix~\ref{app:recovery:parity} we detail the repeated oscillator-parity extraction circuit and bound its majority-vote failure rate. In Appendix~\ref{app:recovery:checks} we describe the stabilizer measurement schedule and detector graph for the repetition code, and in Appendix~\ref{app:recovery:decoder} we bound the resulting logical-frame failure and residual local error probabilities.

Each sensing cycle consists of encoding, sensing, uncomputation, repeated oscillator-parity extraction, the conditional $Z_1$ correction, oscillator reset, and repeated data-check extraction and correction. The signal acts only during the sensing interval. Ideal control operations are followed by the specified independent qubit $X$ faults; oscillator reset prepares vacuum exactly. We allow the full number-dephasing channel during sensing, and number dephasing during parity extraction does not affect the argument below. Classical majority decisions, decoding and feedforward are taken to be reliable.

\subsection{Displacement faults and the oscillator branches}
\label{app:recovery:displacements}

In this appendix we consider the syndrome extraction and correction protocols to correct dephasing on the probe and ancillas.

Write $W=\prod_jV_j$, where $V_j=\exp[(\alpha/d)(a^\dag-a)Z_j]$. The identities
\begin{equation}
 V_jX_j=X_jV_j^\dagger,\qquad
 WX_j=X_jWV_j^{-2}
 \label{eq:min:propagation}
\end{equation}
express propagation of an $X_j$ through an ideal controlled displacement. They do not replace the gate-output fault model with a reversed or imperfect gate. In particular, an output $X_j$ leaves the oscillator unchanged at that instant but changes the control of a subsequent displacement. No other data qubit acquires an $X$ from this propagation.

At zero signal, condition on any classical qubit fault history and any bosonic dephasing Kraus branch. The two data strings remain complementary, say $x$ and $\bar x$. Since their controlled displacements have opposite signs and every dephasing Kraus operator commutes with $P$, the state before parity extraction has the form
\begin{equation}
 c_0\ket\chi\ket x+c_1P\ket\chi\ket{\bar x}.
 \label{eq:min:faultbranches}
\end{equation}
Parity outcome $b$ followed by $Z_1^b$ gives
\begin{equation}
 (-1)^{b x_1}Q_b\ket\chi\otimes
 (c_0\ket x+c_1\ket{\bar x}).
\end{equation}
The factor $(-1)^{b x_1}$ is a common phase. Resetting the oscillator therefore leaves only the data bit pattern to be corrected. The result holds for arbitrary Fock support and requires no expansion in $\alpha/d$.

At finite signal, let $a_{h,j}$ be the real coherent amplitude produced by the faulty encoding in cycle $j$, conditional on the qubit fault history $h$. It is a sum of $d$ signed contributions of magnitude $\alpha/d$, so $|a_{h,j}|\le\alpha$. A single flip before $W$, for example, gives $a_{h,j}=\alpha(1-2/d)$ relative to the original logical labels. A flip after $W$ (before $W^\dag$) leaves the encoded amplitude unchanged. Qubit flips during sensing do not act on the oscillator, but they must be retained in the data error frame.

Faulty uncomputation applies opposite real displacements to the two branches. These displacements cancel from the parity-corrected overlap, since
$D(-u)P D(-u)=P$. More explicitly, if the oscillator states before uncomputation are $U_\theta\ket{a_{h,j}}$ and $U_\theta\ket{-a_{h,j}}$, the recovered coherence is determined by
\begin{equation}
 \langle-a_{h,j}|U_\theta^\dag P U_\theta|a_{h,j}\rangle,
\end{equation}
independently of the real decoding displacement $u$. The finite-signal calculation in Appendix~\ref{app:finite:exact} consequently applies with $\alpha$ replaced by $a_{h,j}$. Correcting a qubit fault need not restore the signal that would have been acquired in that particular cycle.

\subsection{Repeated oscillator-parity extraction}
\label{app:recovery:parity}
\label{app:faults:parity}

Next, we detail the circuit for repeated oscillator-parity measurements and show that majority voting suppresses readout-induced logical phase flips to $O_d(p^r)$.

One extraction uses a fresh readout qubit $R$ initialized in $\ket0_R$, followed by
\begin{equation}
 U_{\rm par}=Q_0\otimes I_R+Q_1\otimes X_R
\end{equation}
and computational-basis measurement. Since $[U_{\rm par},X_R]=0$, preparation or output $X_R$ faults can change the record without changing the parity projection. A useful component circuit is
\begin{equation}
 U_{\rm par}=(I_P\otimes H_R)V_{PR}(I_P\otimes H_R),\qquad
 V_{PR}=I_P\otimes\ket0\bra0_R+P\otimes\ket1\bra1_R.
\end{equation}
It maps $\ket{k}_P\ket0_R$ to $\ket{k}_P\ket{k\bmod2}_R$.

An $X_R$ before the first Hadamard reverses the measured record. An $X_R$ immediately after that Hadamard acts as a scalar on the readout eigenstate. An $X_R$ after $V_{PR}$ can be moved backwards through that gate at the cost of $P$ on the oscillator; within an observed parity sector, $PQ_b=(-1)^bQ_b$ is a scalar. Faults after the final Hadamard or at measurement flip the record. Thus the measurement Kraus operators, conditional on component faults, are the intended parity projectors up to scalar phases and relabelled outcomes. These faults do not reveal or alter the logical amplitudes within a parity sector.

Number dephasing commutes with $P$, and the signal is off during extraction, so successive measurements refer to one fixed parity bit. Use $d$ independent preparations and readouts, and take their majority. Fresh qubits may be supplied sequentially, or one physical readout qubit may be reset between rounds under the same independent preparation-fault model. If one record is wrong with probability at most $q\le c p$, then
\begin{equation}
 p_{\rm par}\le\sum_{k=r}^{d}\binom{d}{k}q^k(1-q)^{d-k}
 =O_d(p^r),\qquad r=(d+1)/2.
 \label{eq:min:parityfail}
\end{equation}
A wrong majority causes a logical $Z_L$ error. An $X$ after the corrective $Z_1$ is instead an ordinary data fault, included in the following data recovery. Qubit idles during extraction and oscillator reset are also included in that recovery.

\subsection{Data checks and the syndrome record}
\label{app:recovery:checks}

We now describe the fault-tolerant extraction of the repetition-code stabilizer checks $Z_jZ_{j+1}$ and the construction of the time-ordered syndrome record across cycles.

To measure $Z_jZ_{j+1}$, initialize a fresh target $R$ in $\ket0$, apply $\operatorname{CNOT}_{j\to R}$ and $\operatorname{CNOT}_{j+1\to R}$, and measure $Z_R$. Both CNOTs use data controls. A target $X_R$ cannot propagate backwards to either data qubit. A data $X_j$ can change the record and persists on that same qubit, but cannot spread to its neighbour through this circuit. Gate-output faults on the two qubits are separate elementary fault locations, as specified by the independent single-qubit model.

Schedule the checks in a fixed order, repeat a full round $d$ times per recovery, and include all intervening idle locations. Write the recorded syndrome bit as $s_{j,t}$, with $0$ for a $+1$ result. The detection events are
\begin{equation}
 D_{j,t}=s_{j,t}\oplus s_{j,t-1}.
\end{equation}
A record error changes consecutive detection events at one check, whereas a data error changes the adjacent checks. A fault occurring part-way through a round can connect events at different times; these circuit-dependent edges must be included. The detector graph is therefore constructed from the actual gate schedule, rather than from a model of perfect simultaneous check measurements.

Use the repetition-memory decoder of Ref.~\cite{conlon2026fault}, Appendix B.6, with minimum fault weight and the appropriate preparation and terminal boundaries. Retain the syndrome history across sensing cycles; do not introduce an assumed perfect syndrome at each cycle boundary. The decoder identifies a data correction modulo the logical operator $X_L$. Apply its single-qubit $X$ corrections, or retain the equivalent Pauli frame with all subsequent controls and reported signs adjusted accordingly. In the protocol described here the corrections are applied physically, and their output faults are included in the next round.

\subsection{Logical failures and residual local errors}
\label{app:recovery:decoder}

Finally, we bound the probability of a decoded logical-frame failure over $M$ cycles and distinguish it from correctable local residual faults that alter the single-cycle encoding amplitude.

The imported memory guarantee is a circuit-distance statement: a nontrivial logical chain has at least $d$ elementary faults or correction edges, with the preparation and terminal treatment specified above. If an actual fault chain and a minimum-fault-weight correction differ by such a chain, at least $r=(d+1)/2$ of its edges must be actual faults. The bounded-degree detector graph then gives, at fixed $d$ and sufficiently small $p$,
\begin{equation}
 \Pr(\text{decoded logical-frame failure})
 \le A_d(M+1)p^r.
 \label{eq:min:memoryfail}
\end{equation}
The factor $M+1$ counts possible positions of a failing component, including preparation. One obtains the bound by summing connected paths of length at least $d$, each containing at least half as many actual faults. Their number grows at most geometrically with length, so the sum converges for sufficiently small $p$. This estimate does not require statistical independence of decoded failures in different cycles.

The guarantee does not imply a fault-free output after every recovery. A late fault can remain as a correctable error until later checks resolve it. For any specified encoding, an incorrect local control pattern outside a decoded logical failure requires a fault-containing connected component meeting that encoding's neighbourhood. Summing these components gives a bound $B_dp$ on the probability of a local gain disturbance. The neighbourhood includes the preceding recovery and the current controlled displacements; components extending further are included in the same convergent path sum. Unresolved local errors at the terminal boundary contribute $O_d(p)$ to the final measurement sign and are counted separately from Eq.~\eqref{eq:min:memoryfail}.

These two estimates have different roles. A logical failure can reverse the interpretation of previously accumulated phase, whereas a local residual error can change one cycle's gain without causing a logical failure. Both contributions must therefore be retained in the Fisher-information calculation. The distance estimate assumes the stated circuit-level decoder and boundaries; applying majority votes separately to the data checks would not supply it.

\section{Final Measurement and Estimation}
\label{app:meas}

In this appendix we describe the terminal readout of the logical register and the resulting parameter estimation procedure after all $M$ sensing cycles are complete. In Appendix~\ref{app:meas:readout} we analyze the transversal logical $Y_L$ measurement and bound the one-time readout contrast loss from terminal faults. In Appendix~\ref{app:meas:estimation} we derive the binary outcome likelihood and construct an explicit locally unbiased estimator for $\theta$.

\subsection{Transversal logical readout}
\label{app:meas:readout}

We first describe the transversal measurement of the logical operator $Y_L$ at the end of the protocol and show that terminal single-qubit faults introduce only an $O_d(p)$ contrast reduction.

The last recovery resets the oscillator, so the final measurement acts only on the retained register. On the repetition code,
\begin{equation}
 Y_L=Y_1X_2\cdots X_d.
\end{equation}
Measure $Y_1$ by applying $S^\dagger$ and then $H$ before computational-basis readout, where $S=\operatorname{diag}(1,i)$. Apply $H$ before readout on every other qubit to measure $X_j$. The product of the outcomes is the reported logical result. If a data correction $X^u$ remains in the decoder's frame, multiply this result by $(-1)^{u_1}$, since only the factor $Y_1$ anticommutes with a data $X$. Known logical $Z$ frame updates similarly reverse the reported sign. Earlier oscillator-parity corrections are applied physically in the stated protocol.

All terminal quantum operations are single-qubit gates. A fault cannot spread to other data qubits and can at worst reverse the final reported product. Counting one $S^\dagger$, $d$ Hadamards and $d$ binary measurements gives
\begin{equation}
 \epsilon_{\rm readout}\le(2d+1)p.
\end{equation}
Terminal idles and unresolved correctable errors at the final decoding boundary increase this to $\epsilon_{\rm end}\le c_{{\rm end},d}p$. This is a one-time loss of contrast, not a claim of $O_d(p^r)$ suppression of the final measurement error. It suffices for the sensing bound because the measurement is performed once, after phase accumulation has finished.

\subsection{Likelihood and a local estimator}
\label{app:meas:estimation}

We now connect the fault-averaged readout expectation value to the classical Fisher information and construct a locally unbiased estimator around the operating point $\theta = 0$.

Let $y(\theta)$ denote the mean of the final, frame-adjusted binary result $z=\pm1$, after averaging over circuit faults and the parity outcomes used for correction. Then
\begin{equation}
 p(z\mid\theta)=\frac{1+z y(\theta)}2,\qquad
 \mathcal F_C(\theta)=\frac{y'(\theta)^2}{1-y(\theta)^2}.
 \label{eq:meas:likelihood}
\end{equation}
For ideal controls and simultaneous dephasing, $y(\theta)=-\operatorname{Im}\chi_\tau(\theta)^M$, with $\chi_\tau$ from Appendix~\ref{app:finite:exact}. For faulty controls the appropriate mean is the fault-history average derived in Appendix~\ref{app:ft:likelihood}; replacing it by the ideal fringe would generally give a miscalibrated estimator.

At the working point, $y(0)=0$. If the actual slope $g=y'(0)>0$ is known from the noise model or calibration, $\nu$ independent complete experiments give the estimator
\begin{equation}
 \hat\theta=\frac{1}{\nu g}\sum_{\ell=1}^{\nu}z_\ell.
\end{equation}
It satisfies $\mathbb E_0\hat\theta=0$, $\partial_\theta\mathbb E_\theta\hat\theta|_0=1$, and
\begin{equation}
 \operatorname{Var}_0(\hat\theta)=\frac{1}{\nu g^2}
 =\frac{1}{\nu\mathcal F_C(0)}.
\end{equation}
Thus the local information bound refers to an explicit measurement and a locally unbiased estimator. A lower bound on $g$ is not itself a calibrated value for the estimator. At a finite residual signal one may instead use the full likelihood in Eq.~\eqref{eq:meas:likelihood}. Global phase localization and any calibration cost are separate resources from the local experiment analysed here.

\section{Fault-tolerant CFI with endpoint dephasing}
\label{app:ftend}

In this appendix we derive the classical Fisher information for the full fault-tolerant sensing protocol when bosonic dephasing acts immediately after the signal unitary in each cycle. In Appendix~\ref{app:ftend:cycle} we evaluate the single-cycle logical coherence multiplier conditional on the circuit fault history. In Appendix~\ref{app:ftend:likelihood} we obtain the exact fault-averaged outcome likelihood and classical Fisher information. In Appendix~\ref{app:ftend:budget} we combine the preparation, parity-readout, and memory error budgets to establish a uniform lower bound on the local Fisher information, which we reformulate as a per-cycle survival bound in Appendix~\ref{app:ftend:survival}. Finally, we compare this with the closed-form result when only parity readout is faulty in Appendix~\ref{app:ftend:parity-only}, and characterize the guaranteed interrogation window and finite-signal regime in Appendix~\ref{app:ftend:window}.

We first analyse the fault-tolerant sensing protocol when the bosonic dephasing channel acts after the signal unitary in
each cycle. The oscillator begins in vacuum, the register controls
the encoding displacements, and the oscillator evolves under
$U_\theta=\exp[-i\theta\tau(a+a^\dagger)]=D(-i\theta\tau)$.
This is followed by a trace-preserving number-dephasing channel
$\mathcal E(\rho)=\sum_\mu K_\mu\rho K_\mu^\dagger$ satisfying
\begin{equation}
 [K_\mu,P]=0,\qquad \sum_\mu K_\mu^\dagger K_\mu=I,
 \qquad P=(-1)^n.
 \label{eq:ftend:kraus}
\end{equation}
Uncomputation, parity measurement, the parity-conditioned $Z_1$
correction, and oscillator reset complete the bosonic part of the
cycle. Reset and classical processing are ideal. We use the
parameter-independent stochastic qubit-$X$ fault model and the
repetition-memory estimates assumed for the manuscript's circuit.
In particular, faults are sampled independently of the ideal parity
outcomes. The parity records are discarded after their correction
purpose has been served. The CFI below refers to the final binary
logical measurement, with its sign adjusted using the data-error frame.

\subsection{One recovered cycle conditional on the circuit faults}
\label{app:ftend:cycle}

We first compute the reduced logical channel for a single sensing and recovery cycle under endpoint dephasing, conditioned on a fixed circuit fault history.

Write $\alpha=\sqrt N$. Conditional on the qubit fault history,
encoding prepares two opposite coherent amplitudes $a$ and $-a$,
where $a\in\mathbb R$ and $|a|\le\alpha$. The data strings remain
complementary. A correctable control fault can change $a$, and
correcting the data later does not restore the signal lost in that
cycle. Let the two uncomputation displacements be $D(-u)$ and $D(u)$,
with $u\in\mathbb R$, and put $Q_b=[I+(-1)^bP]/2$.

After summing the true parity outcomes and applying their correct
logical phase corrections, the multiplier of the logical
off-diagonal element is
\begin{align}
 \chi_a^{\rm end}(\theta)
 &=\sum_{\mu,b}(-1)^b
 \bra{-a}U_\theta^\dagger K_\mu^\dagger
 D(-u)Q_bD(-u)K_\mu U_\theta\ket a\notag\\
 &=\sum_\mu\bra{-a}U_\theta^\dagger K_\mu^\dagger
 P K_\mu U_\theta\ket a\notag\\
 &=\bra{-a}U_\theta^\dagger P U_\theta\ket a\notag\\
 &=\bra a D(-2i\theta\tau)\ket a
 =\exp[-2\theta^2\tau^2-4ia\theta\tau].
 \label{eq:ftend:onecycle}
\end{align}
Here we used $D(-u)PD(-u)=P$, Eq.~\eqref{eq:ftend:kraus}, and
the coherent-state displacement characteristic function. For
complementary erroneous data strings, using physical $Z_1$ for the
parity correction changes only a common phase relative to this
calculation. The reduced logical channel is therefore the same in
the corresponding data-error frame.

Equation~\eqref{eq:ftend:onecycle} includes all dephasing Kraus
operators and requires no small-dephasing approximation. A wrong
reported parity majority adds a logical $Z$ and multiplies the
coherence by $-1$. It does not alter the common factor
$\exp(-2\theta^2\tau^2)$. Oscillator reset preserves the resulting
reduced logical state and permits the next cycle to begin in vacuum.

\subsection{Exact likelihood and CFI for a specified fault distribution}
\label{app:ftend:likelihood}

We now average over the classical circuit fault histories to obtain the exact binary measurement likelihood and classical Fisher information after $M$ cycles.

Let $h$ specify the classical circuit faults and their induced
decoder decisions, with probability $\pi_h$ independent of $\theta$.
The ideal oscillator-parity outcomes are already summed over in
Eq.~\eqref{eq:ftend:onecycle}; they are not part of $h$.
Logical $X$ errors conjugate accumulated coherences. Accounting
for these conjugations, define the real signed phase gain
\begin{equation}
 G_h=4\tau\sum_{j=1}^M b_{h,j},
 \qquad |b_{h,j}|\le\alpha,
 \qquad |G_h|\le4\sqrt N T.
 \label{eq:ftend:gain}
\end{equation}
The coefficient $b_{h,j}$ is the encoding amplitude in cycle $j$
with the sign with which that cycle's phase reaches the final
logical measurement. Let $s_h\in\{+1,-1\}$ collect seed-phase
errors, wrong parity corrections, and terminal sign errors.
Multiplying the recovered cycle channels gives the conditional mean
\begin{equation}
 y_h(\theta)=s_h V(\theta)\sin(\theta G_h),
 \qquad V(\theta)=e^{-2\theta^2T\tau}.
 \label{eq:ftend:conditionalmean}
\end{equation}
This representation includes histories with decoded logical failures;
they can change both the signs of earlier phase contributions and
the subsequent encoding amplitudes.

Introduce the two fault averages
\begin{align}
 S_M(\theta)&=\sum_h\pi_h s_h\sin(\theta G_h),
 \label{eq:ftend:average-sine}\\
 C_M(\theta)&=\sum_h\pi_h s_h G_h\cos(\theta G_h)
 =S_M'(\theta).
 \label{eq:ftend:average-cosine}
\end{align}
The binary outcome $z=\pm1$ has probabilities
\begin{equation}
 p(z\mid\theta)=\frac{1+zV(\theta)S_M(\theta)}2.
 \label{eq:ftend:probabilities}
\end{equation}
At regular points of this likelihood, its CFI is exactly
\begin{equation}
 \mathcal F_C^{\rm end}(\theta)=
 \frac{e^{-4\theta^2T\tau}
 [C_M(\theta)-4\theta T\tau S_M(\theta)]^2}
 {1-e^{-4\theta^2T\tau}S_M(\theta)^2}.
 \label{eq:ftend:exact-cfi}
\end{equation}
In particular, $S_M(0)=0$ and
\begin{equation}
 \mathcal F_C^{\rm end}(0)
 =\left(\sum_h\pi_h s_hG_h\right)^2
 =16\tau^2\left(\sum_{j=1}^M
 \mathbb E_h[s_h b_{h,j}]\right)^2.
 \label{eq:ftend:exact-local}
\end{equation}
Thus the relevant quantity is the square of the average signed
gain. Averaging the individual histories' Fisher informations would
instead assume access to the unobserved physical fault history and
would generally give a different, larger information.

Equations~\eqref{eq:ftend:exact-cfi} and
\eqref{eq:ftend:exact-local} are exact once $\pi_h$ and the circuit
are specified. Bounds on elementary fault probabilities alone do
not determine a unique value of these averages. They do, however,
give the uniform lower bound derived next.

\subsection{Fault budget and the guaranteed local information}
\label{app:ftend:budget}

Next, we combine the error bounds from verified preparation, repeated parity readout, and repetition-code decoding to establish a lower bound on the local classical Fisher information at $\theta = 0$.

Let $r=(d+1)/2$. The preparation and repetition-memory analysis
provides a seed-sign error bound $\epsilon_{\rm seed}=O_d(p^r)$,
a terminal sign-error bound $\epsilon_{\rm end}=O_d(p)$, and
a decoded logical-frame failure bound
$A_d(M+1)p^r$. Outside a decoded logical failure, a local error
disturbs any specified encoding with probability at most $B_dp$.
The latter is a per-encoding bound, not the probability that the
entire run contains no local fault.

Define $p_{\rm par}$ to be the wrong-majority probability in a
parity extraction cycle, or a uniform upper bound on it if cycles
have different error probabilities. With independent records of
identical error probability $q$,
\begin{equation}
 p_{\rm par}=\sum_{k=r}^{d}\binom{d}{k}q^k(1-q)^{d-k}.
 \label{eq:ftend:parity-probability}
\end{equation}
If the independent record-error probabilities are only bounded by
$q$, this expression is an upper bound. Since $q\le c_{\rm par}p$
for the fixed measurement circuit, $p_{\rm par}=O_d(p^r)$.
For example, three identical readings give
$p_{\rm par}=3q^2-2q^3$. This error probability is distinct from
the probability that the oscillator actually has odd parity.

A convenient total bound is
\begin{equation}
 \eta_M=\epsilon_{\rm seed}+\epsilon_{\rm end}
 +B_dp+A_d(M+1)p^r+M p_{\rm par}
 \le b_dp+a_d(M+1)p^r.
 \label{eq:ftend:eta}
\end{equation}
Here $a_d,b_d$ absorb the fixed circuit constants. For each cycle
$j$, outside its local disturbance and the global adverse events
listed in Eq.~\eqref{eq:ftend:eta}, one has
$s_hb_{h,j}=\alpha$. In every history
$-\alpha\le s_hb_{h,j}\le\alpha$. A union bound therefore gives
\begin{equation}
 \mathbb E_h[s_hb_{h,j}]
 \ge\alpha(1-2\eta_M).
 \label{eq:ftend:mean-gain}
\end{equation}
Using Eq.~\eqref{eq:ftend:exact-local}, for $\eta_M<1/2$ we obtain
\begin{equation}
 \mathcal F_C^{\rm end}(0)
 \ge16NT^2(1-2\eta_M)^2.
 \label{eq:ftend:local-bound}
\end{equation}
The order-$p$ term produces a fractional change in the total slope;
it need not be multiplied by $M$. The order-$Mp^r$ terms instead
bound failures that can corrupt the interpretation of the accumulated
phase. The estimates in this subsection use the repetition-memory
guarantees established for the adopted circuit and decoder; the
bosonic channel calculation does not independently establish those
memory guarantees.

\subsection{A per-cycle survival bound}
\label{app:ftend:survival}

We now present an alternative formulation of the local Fisher information bound in terms of the conditional per-cycle survival probability of the logical memory.

The union bound in Eq.~\eqref{eq:ftend:eta} suffices for the scaling
claim but hides the multiplicative survival factor. Suppose that,
conditional on no previous decoded logical failure, the probability
of a new logical failure in each cycle is at most $p_L$, and that
the probability of adverse boundary events conditional on survival
is at most $\epsilon_{\rm bd}$. Independence suffices for these
conditions but is not required. The chain rule then implies
\begin{equation}
 P_{\rm good}\ge(1-\epsilon_{\rm bd})(1-p_L)^M.
 \label{eq:ftend:survival}
\end{equation}
If the probability of a local sensing-gain disturbance in any
specified cycle is at most $B_dp$, the bounded-history argument of
Eq.~\eqref{eq:ftend:mean-gain} gives, whenever the bracket is positive,
\begin{equation}
 \mathcal F_C^{\rm end}(0)\ge16NT^2
 \left[2(1-\epsilon_{\rm bd})(1-p_L)^M-1-2B_dp\right]^2.
 \label{eq:ftend:survival-cfi}
\end{equation}
Here $p_L$ includes wrong parity majorities unless they are treated
separately. The existing aggregate $A_d(M+1)p^r$ failure bound alone
does not prove the conditional per-cycle bound: decoding failures
may be correlated across cycles. At leading order,
$(1-p_L)^M=1-Mp_L+O(M^2p_L^2)$, so both bounds support
$M=O(p^{-r})$. The factor $(1-p_L)^M$ is a no-failure probability;
the square bracket converts it to a worst-case contrast. If every
logical failure acts as an independent symmetric sign flip, the
exact contrast instead is $(1-2p_L)^M$.

\subsection{Closed form when only parity reporting is faulty}
\label{app:ftend:parity-only}

To connect directly with the unprotected protocol in Appendix~\ref{app:faults}, we evaluate the exact classical Fisher information in the simplified limit where only the repeated parity readout experiences faults.

For comparison, suppose that all other operations are ideal and
wrong parity majorities occur independently between cycles with
probability $p_{\rm par}$. There are then no gain disturbances or
logical $X$ failures, so $G_h=4\sqrt N T$ in every history and
$\mathbb E_h s_h=(1-2p_{\rm par})^M$. Defining
$\Lambda_M=(1-2p_{\rm par})^M$ and
$\Phi=4\sqrt N\theta T$, Eq.~\eqref{eq:ftend:exact-cfi} becomes
\begin{equation}
 \mathcal F_{C,\rm parity}^{\rm end}(\theta)
 =\frac{\Lambda_M^2e^{-4\theta^2T\tau}
 [4\sqrt N T\cos\Phi-4\theta T\tau\sin\Phi]^2}
 {1-\Lambda_M^2e^{-4\theta^2T\tau}\sin^2\Phi}.
 \label{eq:ftend:parity-cfi}
\end{equation}
Consequently,
\begin{equation}
 \mathcal F_{C,\rm parity}^{\rm end}(0)
 =16NT^2(1-2p_{\rm par})^{2M}.
 \label{eq:ftend:parity-local}
\end{equation}
Without repeated parity readout, replace $p_{\rm par}$ by the
single-record error probability $q$. The resulting expression is
the unprotected comparison in the main text. Merely making this
replacement in the opposite direction does not give the exact CFI
of the full faulty circuit: data errors, seed errors and terminal
faults also enter Eqs.~\eqref{eq:ftend:exact-cfi}--\eqref{eq:ftend:eta}.

\subsection{Useful interrogation times and finite residual signals}
\label{app:ftend:window}

Finally, we determine the maximum interrogation window over which Heisenberg scaling is guaranteed and extend our lower bound on the Fisher information to a finite interval of nonzero signals $\theta$.

At $\theta=0$, define a useful run by retaining a chosen fixed
fraction $0<\rho<1$ of the ideal CFI $16NT^2$.
For unprotected parity reporting with $0<q<1/2$, the exact condition is
\begin{equation}
 M\le\frac{\log\rho}{2\log(1-2q)}
 =\frac{-\log\rho}{4q}[1+O(q)].
 \label{eq:ftend:bare-window}
\end{equation}
For the full fault-tolerant protocol, a sufficient condition is
\begin{align}
 \eta_M&\le\eta_\rho:=\frac{1-\sqrt\rho}{2},\notag\\
 M&\le\frac{\eta_\rho-b_dp}{a_dp^r}-1,
 \qquad T=M\tau,
 \label{eq:ftend:ft-window}
\end{align}
provided the numerator is positive and the fixed-distance memory
bounds apply. An integer number of cycles is chosen below these
bounds. For example, retaining half the ideal CFI gives
$M\lesssim0.1733/q$ in the unprotected parity-only model and is
guaranteed for the full FT protocol when
$b_dp+a_d(M+1)p^r\le0.1464$.
The latter is a sufficient guarantee, not an exact maximum usable
time or an optimum over all protocols. Absolute numerical windows
require the circuit-dependent constants. Taking $q$ proportional to
$p$ gives the comparison $T=O(\tau p^{-1})$ versus a guaranteed
window of order $\tau p^{-r}$. An upper bound $q\le c_{\rm par}p$
alone does not establish that the unprotected window is only of
order $\tau p^{-1}$.

For completeness, the common envelope in the endpoint model also
gives a finite-signal bound. Put $G_{\max}=4\sqrt N T$. Using
$|1-\cos x|\le x^2/2$ and $|\sin x|\le|x|$,
\begin{align}
 C_M(\theta)&\ge C_M(0)-\tfrac12\theta^2G_{\max}^3,\notag\\
 |S_M(\theta)|&\le |\theta|G_{\max}.
 \label{eq:ftend:derivative-control}
\end{align}
Together with $C_M(0)\ge G_{\max}(1-2\eta_M)$ and
$\mathcal F_C\ge [y'(\theta)]^2$ at regular points, this yields
\begin{equation}
 \mathcal F_C^{\rm end}(\theta)
 \ge16NT^2 e^{-4\theta^2T\tau}
 \left[1-2\eta_M-\theta^2(8NT^2+4T\tau)\right]_+^2,
 \label{eq:ftend:finite-bound}
\end{equation}
where $[x]_+=\max\{x,0\}$. For fixed $N>0$ and
$\eta_M\le\eta_*<1/2$, the interval
$|\theta|\le c_\theta/(\sqrt N T)$ with sufficiently small
$c_\theta>0$ retains an $NT^2$ lower bound with a nonzero
constant prefactor. The constant can depend on $N$ and $\eta_*$;
for $M\ge1$, the correction in the bracket is at most
$c_\theta^2(8+4/N)$. Thus the guaranteed interval of residual
signals shrinks with the run length. The result does not assert
uniform Heisenberg scaling at fixed nonzero $\theta$ as $T$ grows.

All times above are sensing times. If a cycle takes elapsed time
$\tau_{\rm cyc}$, the elapsed run time is $M\tau_{\rm cyc}$.
When $p\simeq\gamma_A\tau_{\rm loc}$ at fixed operation durations,
the guaranteed sensing window scales as
$\tau(\gamma_A\tau_{\rm loc})^{-r}$. Uncorrected noise channels,
including qubit phase faults, impose additional restrictions beyond
the bit-flip model analyzed here.

\section{Fault-Tolerant Sensing Bound}
\label{app:ft}
\label{app:ft:model}
\label{app:ft:theorem}

In this appendix we establish the fault-tolerant sensing bound in the realistic regime where photon-number dephasing acts continuously throughout each sensing interval. In Appendix~\ref{app:ft:likelihood} we combine the stochastic phase-diffusion trajectories from Appendix~\ref{app:finite:exact} with the circuit fault histories to lower-bound the local classical Fisher information. In Appendix~\ref{app:ft:finite} we extend this bound to a finite interval of unknown residual signals, and in Appendix~\ref{app:ft:resources} we summarize the resulting interrogation time scaling, duty cycle overhead, and sensitivity to uncorrected ancilla phase faults.

\subsection{Fault histories and the logical likelihood}
\label{app:ft:likelihood}

We now consider the fault-tolerant sensing bound when the dephasing acts throughout the sensing window. This represents a more realistic setting.

Average the final measurement probabilities over circuit faults before evaluating Fisher information. Let $h$ specify all classical circuit faults, but not the ideal oscillator-parity outcomes. The latter are summed over as part of the recovery channel. This distinction matters because the oscillator-parity probabilities themselves can depend on $\theta$. Wrong-majority events in Eq.~\eqref{eq:min:parityfail} are nevertheless parameter independent: the independent record flips are defined relative to the actual parity bit.

Conditional on $h$, the parity/reset calculation extends Eq.~\eqref{eq:min:chi} to each sensing cycle, with a real encoding amplitude $a_{h,j}$ satisfying $|a_{h,j}|\le\alpha$. Logical $X$ errors conjugate previously acquired coherences, and erroneous parity corrections multiply them by $-1$. Accordingly the final logical-$Y$ mean can be written
\begin{equation}
 y(\theta)=\mathbb E\left[s_h e^{-2\theta^2 R}
                                  \sin(\theta S)\right],
 \quad
 R=\sum_{j=1}^M|A_j|^2\le T\tau,\quad
 S=4\sum_{j=1}^M b_{h,j}\operatorname{Re}A_j,
 \quad |b_{h,j}|\le\sqrt N.
 \label{eq:min:history}
\end{equation}
The expectation includes independent phase diffusion in each sensing interval. In the endpoint-dephasing model, set $A_j=\tau$ deterministically, so $R=T\tau$ and the formulas below hold with $f=1$. The sign $s_h$ contains wrong parity corrections, seed signs and final outcome flips. The coefficients $b_{h,j}$ contain the encoding amplitudes and the transported logical frame. Their dependence on faults is parameter independent, and $|S|\le4\sqrt N T$. The formula follows by multiplying the one-cycle coherence factors, conjugating them whenever a logical $X$ changes the frame, and summing the parity outcomes before tracing out each oscillator. Decode displacements cancel from each branch overlap, as established in Appendix~\ref{app:recovery:displacements}.

Let $\epsilon_{\rm seed}$ and $\epsilon_{\rm end}$ bound seed-phase and terminal readout faults. Combining the memory estimates with oscillator-parity majority failures gives
\begin{equation}
 \eta_M=\epsilon_{\rm seed}+\epsilon_{\rm end}
       +B_dp+A_d(M+1)p^r+M p_{\rm par}.
 \label{eq:min:eta}
\end{equation}
The local term $B_dp$ bounds the disturbance of any specified cycle; it is not the probability that no local fault occurred anywhere in the run. For each cycle, outside its local disturbance and the listed global adverse events the coefficient transported to readout is $s_h b_{h,j}=\alpha$; in every event it lies between $-\alpha$ and $\alpha$. Since the circuit faults are independent of phase diffusion, $\mathbb E\operatorname{Re}A_j=\tau f(\kappa\tau)$ implies
\begin{equation}
 y(0)=0,\qquad y'(0)=\mathbb E(s_h S)
 \ge4\sqrt N T f(\kappa\tau)(1-2\eta_M).
\end{equation}
For $\eta_M<1/2$, the binary likelihood therefore has
\begin{equation}
 \mathcal F_C(0)\ge16NT^2 f(\kappa\tau)^2(1-2\eta_M)^2.
 \label{eq:min:ftlocal}
\end{equation}
The bound follows from the likelihood derivative, not from a state-fidelity estimate.

\subsection{A nonzero interval of unknown signals}
\label{app:ft:finite}

We now consider the fault-tolerant sensing bound when the dephasing acts throughout the sensing window. This represents a more realistic setting.

Differentiating Eq.~\eqref{eq:min:history} and using $1-e^{-u}\le u$, $|1-\cos v|\le v^2/2$ and $|\sin v|\le|v|$ gives
\begin{align}
 |y'(\theta)-y'(0)|
 &\le\theta^2\mathbb E\left[6R|S|+\tfrac12|S|^3\right]\\
 &\le4\sqrt N T\theta^2(6T\tau+8NT^2).
 \label{eq:min:derivativebound}
\end{align}
For a binary mean, $\mathcal F_C(\theta)=y'(\theta)^2/[1-y(\theta)^2]\ge y'(\theta)^2$ at regular points. Therefore
\begin{equation}
\mathcal F_C(\theta)\ge16NT^2
 \left[f(\kappa\tau)(1-2\eta_M)
              -\theta^2(6T\tau+8NT^2)\right]_+^2,
 \label{eq:min:ftfinite}
\end{equation}
where $[x]_+=\max\{x,0\}$. This conservative bound is uniform on a finite interval of unknown residual signals. In particular, $|\theta|\le c/(\sqrt N T)$ with sufficiently small fixed $c$ permits a finite accumulated logical phase and retains $NT^2$ scaling. For $M=a_0p^{-r}$, this sufficient interval is $|\theta|\tau\le c p^r/(a_0\sqrt N)$. The interval shrinks with interrogation time; no assertion of uniform Heisenberg scaling at fixed nonzero $\theta$ and fixed cycle time is made. Larger signals can be studied with the exact likelihood, a shorter cycle or a compensating reference drive. Global localization of an initially unrestricted signal requires a separate estimation procedure.

\subsection{Interrogation window and resources}
\label{app:ft:resources}

Finally, we summarize how the fault-tolerant interrogation window scales with the physical fault rate and code distance, while accounting for cycle latency and uncorrectable ancilla phase errors.

Verified preparation gives $\epsilon_{\rm seed}=O_d(p^r)$, final readout gives $\epsilon_{\rm end}=O_d(p)$, and $p_{\rm par}=O_d(p^r)$. At fixed distance, schedule and sensing interval, sufficiently small $p$ allows $M=\lfloor a p^{-r}\rfloor$ with a sufficiently small positive constant $a$ while keeping the bracket in Eq.~\eqref{eq:min:ftfinite} bounded away from zero in the stated residual-signal range. Thus the fault-tolerant interrogation window is of order $\tau p^{-(d+1)/2}$. If $p$ is proportional to $\gamma_A\tau_{\rm loc}$ at fixed location durations, its physical-rate exponent is $\gamma_A^{-(d+1)/2}$. All distance-dependent prefactors remain in the memory and measurement constants; the argument does not establish a uniform large-distance threshold.

If a cycle takes elapsed time $\tau_{\rm cyc}$, then $T=M\tau$ and the information expressed in elapsed time includes $(\tau/\tau_{\rm cyc})^2$. If rare additional stochastic phase faults are included, a bound $\epsilon_Z$ on their occurrence anywhere in the run can be added to $\eta_M$; independent phase flips satisfy $\epsilon_Z\le\gamma_Z\mathcal T_A$ for total exposed qubit-time $\mathcal T_A$. Repetition coding does not suppress them. The energy constraint is exact throughout sensing at $\theta=0$. At nonzero signal it applies initially; imposing it at every time requires a preparation margin or a separate energy bound.

\twocolumngrid
\bibliography{ECbib}
\end{document}